\documentclass[12pt,onecolumn]{IEEEtran}
\IEEEoverridecommandlockouts
\let\labelindent\relax
\usepackage{cite}
\usepackage{amsmath,amssymb,amsfonts}
\usepackage{graphicx}
\usepackage{textcomp}
\usepackage{xcolor}
\usepackage{epsfig}
\usepackage{amssymb}
\usepackage{amsbsy}
\usepackage{cite}
\usepackage{enumitem}
\usepackage{url}
\usepackage{xcolor}
\usepackage{color}
\usepackage{stackengine}
\usepackage{mathtools}
\usepackage{float}
\usepackage{algorithm}
\usepackage{times,amsmath}
\usepackage{xspace,latexsym,syntonly}
\usepackage{amssymb}
\usepackage{amsfonts}
\usepackage{textcomp}
\usepackage{subcaption}
\usepackage[utf8]{inputenc}
\usepackage{autobreak}
\usepackage{tabularx}
\usepackage{setspace}
\newtheorem{theorem}{Theorem}
\newsavebox\figureone
\newsavebox\figuretwo

\DeclareMathOperator*{\argmin}{arg\,min}
\renewcommand{\baselinestretch}{1.75}

\begin{document}

\title{An Online Learning Approach to Shortest Path and Backpressure Routing in Wireless Networks}

\author{Omer Amar and Kobi Cohen
\thanks{Omer Amar and Kobi Cohen are with the School of Electrical and Computer Engineering, Ben-Gurion University of the Negev, Beer Sheva 8410501 Israel. Email:omerama@post.bgu.ac.il, yakovsec@bgu.ac.il}
\thanks{This work has been submitted to the IEEE for possible publication. Copyright may be transferred without notice, after which this version may
no longer be accessible.}
\thanks{A short version of this paper was presented at the IEEE International Symposium on Information Theory (ISIT) 2021 \cite{amar2021online}.}
\thanks{This research was supported by the ISRAEL SCIENCE FOUNDATION
(grant No. 2640/20), and the Israel Ministry of Economy (Magnet).}
}
\maketitle

\begin{abstract}
We consider the adaptive routing problem in multihop wireless networks. The link states are assumed to be random variables drawn from unknown distributions, independent and identically distributed across links and time. This model has attracted a growing interest recently in cognitive radio networks and adaptive communication systems. In such networks, devices are cognitive in the sense of learning the link states and updating the transmission parameters to allow efficient resource utilization. This model contrasts sharply with the vast literature on routing algorithms that assumed complete knowledge about the link state means. The goal is to design an algorithm that learns online optimal paths for data transmissions to maximize the network throughput while attaining low path cost over flows in the network. We develop a novel Online Learning for Shortest path and Backpressure (OLSB) algorithm to achieve this goal. We analyze the performance of OLSB rigorously, and show that it achieves a logarithmic regret with time, defined as the loss of an algorithm as compared to a genie that has complete knowledge about the link state means. We further evaluate the performance of OLSB numerically via extensive simulations, which support the theoretical findings and demonstrate its high efficiency.\vspace{0.2cm}

\vspace{0.4cm}
\end{abstract}
\IEEEpeerreviewmaketitle

\section{Introduction}
\label{sec:introduction}
Due to the increasing demand of wireless communications along with spectrum scarcity and network dynamics, developing routing algorithms that utilize spectral resources and schedule data transmissions efficiently is a main challenge in communication networks. 
Traditional algorithms assumed complete knowledge about the link state means when scheduling transmissions over selected paths. However, they become inefficient in the era of dynamic networks, adaptive communications and cognitive radio networks, since the link states vary randomly following unknown distributions. Furthermore, the user loads are dynamic and heterogeneous, and need to be balanced. Therefore, in recent years, developing data transmission algorithms based on online learning for adaptive routing in an unknown environment has attracted a growing interest in dynamic networks, distributed learning, adaptive communications and cognitive radio networks \cite{liu2012adaptive,   tehrani2013distributed, pourpeighambar2019joint, scarlett2019overlapping, zhao2020novel, salameh2020intelligent, raj2020survey, gafni2021federated, huang2021tsor}.  

We consider a time-slotted cognitive radio network, where each link state is modeled by a random process drawn from an unknown distribution, independent and identically distributed (i.i.d.) across time and other links, as in \cite{somekh2008cooperative, gai2010learning, liu2012adaptive,   tehrani2013distributed, he2013endhost, ghosh2016secondary, talebi2017stochastic}. The link state represents an effect of the link quality caused by an external process, e.g., by primary users in hierarchical cognitive radio networks, or a fading channel effect in the open sharing cognitive radio model \cite{zhao2007survey}. We define the path state (or path cost) at time slot $t$ as the accumulated states of all links on the path at time slot $t$ (e.g., when summing over path rate measures, delay effects, small packet-drop probability effects \cite{liu2012adaptive, talebi2017stochastic}). 
Once a packet reaches its destination, the random path state of the traveled path is observed by the transmitter (e.g. by ACK signal information \cite{srikant2013communication}). A source-destination pair traffic is denoted by flow, and the set of flows in the network is denoted by $\mathcal{F}$. Flow $f \in \mathcal{F}$ at time slot $t$ generates $A_f(t)$ packets,  with arrival rate $\lambda_f$. The goal is to design a routing and scheduling algorithm for flow transmissions in the network that maximizes the network throughput, while attaining low sum costs over flows in the network (see Section \ref{sec:model} for an explicit formulation). 

\subsection{Routing Algorithms with Complete Knowledge on Link States}
\label{ssec:deterministic}
A well-known approach to routing algorithms is to compute shortest paths for data transmissions. Under complete knowledge of the link states, the shortest path is computed by the minimal accumulated cost over links in a path among all possible paths for data transmissions. For instance, the popular Open Shortest Path First (OSPF) routing protocol uses Dijkstra algorithm to compute shortest paths for data transmissions \cite{srikant2013communication, gong2016distributed}. An alternative approach, dubbed backpressure routing, that routes data in directions that maximize the differential queue backlog between nodes, has attracted a growing attention in recent years since it achieves the maximum network throughput \cite{stolyar2005maximizing, eryilmaz2006joint, bui2009novel, srikant2013communication, sinha2017optimal, joo2011performance}. However, backpressure routing becomes inefficient when the network congestion is low since packets use long paths due to backpressuring data transmissions. As a result, combining shortest path and backpressure routing that uses shorter paths when the network congestion decreases (to avoid large delays by backpressure routing), and longer paths when the network congestion increases (to avoid heavily-loaded links when using shortest path routing) was studied in recent years (see \cite{ying2010combining} and subsequent studies), and shown to maximize the network throughput with low path costs.

\subsection{Online Learning for Adaptive Routing under Unknown Link States}
\label{ssec:stochastic}
In practical adaptive communication systems, the link states are drawn from an unknown distribution, such that their mean values are unknown and need to be learned online. Therefore, recent studies on cognitive radio networks and adaptive communications have focused on developing data transmission algorithms that learn the link states over time and update the transmission parameters to allow efficient resource utilization in the network. Single-hop transmission strategies have been developed using game-theoretic learning \cite{menache2008rate, menache2011network, cohen2015distributed, cohen2016distributed, bistritz2018game, cao2014qos}, multi-armed bandit learning \cite{tekin2011online, tekin2012online, liu2012learning, cohen2014restless, gafni2018learning, bistritz2018distributed, turgay2019exploiting, yemini2020restless, gafni2020learning, gafni2021distributed, gafni2021learning} that often uses reinforcement learning strategies based on Upper Confidence Bound (UCB)-type algorithms \cite{agrawal1995sample, auer2002finite, tabei2021multi}, distributed exploration and exploitation based algorithm \cite{huang2021tsor}, and deep reinforcement learning \cite{wang2018deep, yu2019deep, naparstek2017deep, naparstek2018deep, lin2010autonomic}. Other existing methods for adaptive routing in ad-hoc wireless networks were presented in \cite{zhao2020novel, tang2016joint}. The problem of online learning for adaptive routing in multi-hop transmissions was studied to solve online shortest path routing in \cite{rong2008enhanced, liu2012adaptive,  tehrani2013distributed, he2013endhost,  talebi2017stochastic}. Specifically, the idea in these papers is to develop an efficient learning algorithm by trading-off between exploration of sub-optimal paths and exploitation of the shortest path. In \cite{rong2008enhanced, liu2012adaptive,  tehrani2013distributed, he2013endhost}, the authors focused on making end-to-end route decisions, where in \cite{talebi2017stochastic}, the authors focused on making hop-by-hop decisions, which was shown to be beneficial in adaptive communications due to dynamic route adjustments. In this paper, we allow hop-by-hop decisions as well. Although the algorithms in \cite{rong2008enhanced, liu2012adaptive,  tehrani2013distributed, he2013endhost,  talebi2017stochastic} aim to converge to the shortest path strategy, they do not perform well in terms of load balancing as explained in Subsection \ref{ssec:deterministic}. This issue is particularly relevant in cognitive radio networks, where external primary users might influence the network state and resource usage dynamically with time.

\subsection{Main Results}
We address the adaptive routing problem under unknown link states. Our goal is to design an adaptive routing algorithm that learns online optimal paths for data transmissions to maximize the network throughput, while attaining low path cost over flows in the network. We adopt the routing optimization in \cite{ying2010combining} to achieve this goal. Solutions to the deterministic optimization problem in \cite{ying2010combining} (and variations) have been studied in recent years under complete knowledge of all path state means, as discussed in Subsection \ref{ssec:deterministic}. However, solving the problem in the online learning context without assuming prior knowledge of path state means remained open. This is the first paper to address this problem.

In terms of algorithm development, we develop a novel routing algorithm, dubbed Online Learning for Shortest path and Backpressure (OLSB) algorithm, under unknown link state means. In OLSB, each flow arrived at the source node for transmission is assigned a desired cost to its path. This is done by optimizing a predetermined tradeoff function between the path state and path load, and at the same time learning the unknown path states. In contrast to existing online learning for adaptive routing studies, where the optimal solution considers a single and fixed best path in terms of the expected cost (see e.g., \cite{liu2012adaptive, talebi2017stochastic} and references therein), the path selection of the optimal solution in this paper is time-varying due to the queue dynamics. This leads to fundamentally different design and analysis of the learning algorithm. Specifically, we develop a novel UCB-type rule, dubbed Queue UCB (QUCB), used in the OLSB algorithm. In QUCB, a path selection index that takes into account the dynamic queue state and the path state mean, is developed for adaptive path selections. 
The OLSB algorithm uses the QUCB rule to determine the cost limit for packet transmissions, and backpressures packets through paths that meet the QUCB's cost conditions. The algorithm is described in detail in Section \ref{sec:algorithm}.

In terms of performance analysis, we provide rigorous analysis to evaluate the performance of the OLSB algorithm. To analyze the performance theoretically, our benchmark for performance is defined by a genie that solves the optimization problem with complete knowledge on the link state means, which is known to maximize the network throughput \cite{ying2010combining}. We evaluate the performance of the proposed OLSB algorithm analytically by the regret, defined as the reward loss of OLSB (that operates under unknown link state means) with respect to genie as described above.
As a result, the regret evaluates how fast the proposed OLSB algorithm learns the side information and approaches genie's performance.
We prove analytically that OLSB achieves a logarithmic regret order with time, which indicates that OLSB approaches the performance of genie as time increases with the best known rate. 
Finally, we present extensive simulation results to support the theoretical findings numerically and validate the regret order of the OLSB algorithm. The theoretical and numerical analyses are described in detail in Sections \ref{sec:performance} and \ref{sec:experiemnts}, respectively.

\subsection{Organization}

The rest of this paper is organized as follows: In Section \ref{sec:model}, we present the system model and formulate the problem. In Section \ref{sec:algorithm}, we present the proposed Online Learning for Shortest path and Backpressure (OLSB) algorithm to achieve the objective. In Section \ref{sec:performance}, we analyze the performance of the OLSB algorithm rigorously theoretically, and show that it achieves a logarithmic regret with time. Detailed proofs are given in the Appendix. In Section \ref{sec:experiemnts}, we present simulation results to validate the theoretical findings, and demonstrate the efficiency of the OLSB algorithm. Section \ref{sec:conclusion} concludes the paper. 

\section{System Model and Problem Statement}
\label{sec:model}
We consider a directed graph $G=(V,E)$ where $V$ is the set of nodes and $E$ is the set of edges (or links). Time is slotted, and the time slot index is denoted by $t$. A link from node $v$ to neighbor node $v'$ in $E$ is denoted by $(v,v')$. Each node holds a packet queue for transmissions over links (which will be described later in detail) under a certain MAC protocol. A flow from source node $s \in V$ to destination node $d \in V$ is denoted by $f_{(s,d)}$. We denote the set of all flows in the network by $\mathcal{F}$. We consider the general model of multiple flows that share the network resources. The arrival rate of flow $f_{(s,d)}$ is denoted by $\lambda_{(s,d)}$. 

Every link $e \in E$ is associated with a weight $w_e(t)$ at time slot $t$, which is a random process drawn from an unknown distribution on a normalized support $[0,1]$. The weight $w_e(t)$ is assumed to be i.i.d. across time and other links, as in \cite{somekh2008cooperative, gai2010learning, liu2012adaptive,   tehrani2013distributed, he2013endhost, ghosh2016secondary, talebi2017stochastic}. The set of all possible loop-free paths from any node $v$ to destination $d$ in $G$ is denoted by $\mathcal{P}_{(v,d)}$. The loop-free path from node $v$ to destination node $d$ is denoted by $p\in\mathcal{P}_{(v,d)}$, where $p$ can be represented by a sequence of nodes from $v$ to $d$, e.g., $p=(v, v_1, v_2, ..., v_I, d)$, or either a sequence of links from $v$ to $d$, e.g., $p=\left((v, v_1), (v_1, v_2), ..., (v_I, d)\right)$. The path state (or path cost) $C_p(t)$ for path $p$ at time slot $t$ is defined by the normalized sum of all link weights on that path: $C_p(t) = \frac{1}{|V|}\sum_{\substack{e \in p}}{w_e(t)}$. Note that $0\leq C_p(t)\leq 1$.

As introduced in \cite{ying2010combining}, the objective is to maximize the network throughput (i.e., support the capacity region by using backpressured paths), while attaining low sum costs over flows in the network (by using short paths). Specifically, with complete knowledge of all path state means, $\mu_p = \sum_{e \in p}{E(w_e(t))}, p \in \mathcal{P}_{(s,d)}$, the throughput-optimal solution is to solve the following deterministic optimization problem at each time $t$ \cite{ying2010combining}:\vspace{0.4cm} 
\begin{equation}
    \label{eq:genie}
    \argmin\limits_{p \in \mathcal{B}_{(s,d)}}{\bigg(K\mu_p + Q_{(s,d,m(\mu_{p}))}(t)\bigg)},\vspace{0.4cm}
\end{equation}
where $\mathcal{B}_{(s,d)}$ is a barycentric spanner on the path set $\mathcal{P}_{(s,d)}$ (see Section \ref{ssec:preprocess} for details), and $Q_{(s,d,m(\mu_{p}))}(t)$ is the number of packets (i.e., queue state) in the $m(\mu_{p})$th queue of node $s$ destined to node $d$ by time $t$, where $m(\mu_{p})$ is a mapping function from $\mu_{p}$ to a queue index stored by the node. The term $K$ is a tuning parameter used to balance between short paths and backpressured paths. 
Intuitively, the solution tends to use short paths when the network congestion is light, and backpressured long paths when the network congestion increases. 

Solutions to the deterministic optimization problem \eqref{eq:genie} and variations have been studied in recent years under complete knowledge of all path state means (see \cite{ying2010combining} and subsequent studies). However, solving the problem in the online learning context without assuming prior knowledge of path state means remained open. In this paper we address this problem. The objective of this paper is thus to develop an algorithm that converges (the performance measure is described later) to the solution of \eqref{eq:genie} in the online learning context under unknown path states. We are thus facing an online learning problem with the well-known exploration versus exploitation dilemma. On the one hand, the algorithm should explore all paths in order to infer their states. On the other hand, it should exploit the information gathered so far to route packets in the optimal paths (which vary at each given time). The performance of online learning algorithms are
commonly evaluated by the regret, defined as the loss of
an algorithm as compared to genie with side information
on the system. Here, we wish to design an algorithm that minimizes the regret with respect to the optimal solution of \eqref{eq:genie} (i.e., with complete knowledge of all path state means). In Section \ref{sec:algorithm}, we develop the OLSB algorithm to solve this problem. In Section \ref{sec:performance}, we analyze the performance of OLSB rigorously and prove analytically that it achieves a logarithmic regret order with time, which indicates that it approaches the performance of genie as time increases with the best known rate. 

\section{The Online Learning for Shortest Path and Backpressure (OLSB) algorithm}
\label{sec:algorithm}
In this section, we present the OLSB algorithm to achieve the objective. Different from shortest path-type routing that allows to route packets through a single path (the shortest one) and backpressure routing that allows to route packets in very long paths, OLSB selects a path among all paths with cost less than a path cost constraint determined by the algorithm to tradeoff between short paths and backpressured paths. Specifically, let $C_0, C_1, ..., C_M$ be $M+1$ values, such that $0=C_0<C_1<\cdots< C_{M-1}<1<C_M$. These values are used to quantize the path cost in the network (e.g., distributed with equal intervals). Each node $v\in V$ holds $M$ packet queues for each destination node, corresponding to path constraints with values $C_0, C_1, ..., C_{M-1}$.

Let $m(c): [0, 1]\to \{0,...,M-1\}$ be a mapping function from a cost to a quantized cost level, such that $m(c)=i$ iff $C_i\leq c< C_{i+1}$ ($0\leq i\leq M-1$). When a packet directed to node $d$ with path constraint $c$, such that $C_i\leq c< C_{i+1}$ ($0\leq i\leq M-1$), arrives at node $v$, then node $v$ enters the packet to one of its queues $0, 1, 2, ..., m(c)=i$ (corresponding to $C_0, C_1, ..., C_i$, respectively) destined to node $d$. The queue selection is done by solving a stochastic optimization defined by the OLSB algorithm as will be described later. If node $v$ enters the packet to the $j$th queue ($0\leq j\leq i$), then the path constraint for the packet is updated to $C_j$. Packets in queue $j=1, ..., i$ are delivered to destination via backpressured paths with cost less than $C_j$ only (i.e., the algorithm trades off between backpressured and short paths as will be described in detail later). Packets in queue $0$ (corresponding to path cost $C_0=0$) are delivered through the shortest path only. We denote the queue state $Q_{(v,d,m)}(t)$ as the number of packets in the $m$th queue of node $v$ destined to node $d$ by time $t$. 

Next, we detail the OLSB algorithm using three main phases. The pseudocode of OLSB is given in Algorithm \ref{alg:olsb}. 

\subsection{Preprocessing}
\label{ssec:preprocess}
As commonly done in adaptive routing (see e.g., \cite{awerbuch2008online, liu2012adaptive} and subsequent studies), OLSB uses dependencies between paths to reduce the number of paths that nodes learn by performing a barycentric spanner. For node $v$ and destination node $d$ we apply a barycentric spanner on the paths set $\mathcal{P}_{(v,d)}$ to get a smaller barycentric path set $\mathcal{B}_{(v,d)}$. 

Since the link states and path cost to destinations are random processes with unknown distributions, in OLSB, each node $v$ computes an estimate of the path cost mean $\bar{C}_p(t)$ for all $p\in \mathcal{B}_{(v,d)}$ for each destination node $d$. We define $T_p(t)$ as the number of times path $p\in \mathcal{B}_{(v,d)}$ was selected for transmission after $t$ time steps. This information will be used for efficient learning of the path states. 

In the initialization step, each destination node $d$ transmits one packet through every possible path $p \in \mathcal{B}_{(s,d)}$ for each flow $(s,d) \in \mathcal{F}$. Let $p=(s, v_1, v_2, ..., v_I, d)\in \mathcal{B}_{(s,d)}$, and $p_i=(v_i, v_{i+1}, ..., v_I, d)$ be a sub-path of $p$. The random path cost $C_p(0)$ of path $p$ is observed at the source node, and the random path cost $C_{p_i}(0)$ of any sub-path $p_i$ is observed at node $v_i$ (e.g., each link adds its random weight to the message and any sub-path cost is observed through the path). We set the estimate path cost mean  $\bar{C}_{p}(0)=C_p(0)$ for path $p$ at source node $s$, and $\bar{C}_{p_i}(0)=C_{p_i}(0)$ for sub-path $p_i$ at node $v_i$. We set $T_p(0)=1, \forall p \in \mathcal{B}_{(s,d)}$. 

During the routine of OLSB algorithm, the mechanism described above is implemented using ACK signals from the destination to the source node whenever a packet (or a frame of packets) is delivered through each path in the barycentric spanner\footnote{Paths which are not in the barycentric spanner can be revealed by a simple linear combination of paths in the barycentric spanner \cite{liu2012adaptive}.}. The estimate path cost mean $\bar{C}_{p}(t)$ for path $p$ at source node $s$, and $\bar{C}_{p_i}(t)$ for sub-path $p_i$ at node $v_i$ are computed by the empirical mean for each path $p \in \mathcal{B}_{(s,d)}$ for each flow $(s,d) \in \mathcal{F}$. We note that each path state is evaluated only in one direction, the transmission to the destination, and not by the ACK signal returning to the source node.

\subsection{Packet arrival at the source node}
\label{ssec:packets arrival}
We next describe the algorithm operation at time slot $t$. Consider a packet (or a frame of packets) of flow $f_{(s, d)}$ that arrives to source node $s$ for transmission to destination node $d$. Upon the packet arrival, OLSB needs to select queue $Q_{(s,d,m)}$ ($1\leq m\leq M$) for packet injection among the $M$ packet queues of flow $f_{(s, d)}$. Intuitively, the selection is based on the queue states and the estimated path costs. The priority of injecting a packet to a certain queue increases as the queue load and the path cost constraint decrease. 

To solve the online path selection problem under unknown link states and link distributions, we develop a novel UCB-type online learning rule, which takes into consideration the time-varying queue states and the estimated costs in the online decision making, dubbed Queue UCB (QUCB). 
Specifically, at time slot $t$, packets in flow $f_{(s, d)}$ are injected to the $m(\bar{C}_{p^*}(t))$th queue, i.e., $Q_{(s,d,m(\bar{C}_{p^*}(t)))}$, where $p^*$ is the solution to the following QUCB's stochastic optimization problem:\vspace{0.4cm}
\begin{equation}
    \label{eq:qucb optimization}
    p^* = \argmin\limits_{p \in \mathcal{B}_{(s,d)}}{\bigg(K\bar{C}_p(t) + Q_{(s,d,m(\bar{C}_p(t)))}(t) - \sqrt{\frac{2\ln{t}}{T_p(t)}}\bigg)}, \vspace{0.4cm}
\end{equation}
where the parameter $K$ is a design parameter used to balance between short paths and backpressured paths.
It was shown in \cite{ying2010combining} that when the path state means (say $\mu_p$ of path $p$) are known, the following deterministic optimization: $\argmin\limits_{p \in \mathcal{B}_{(s,d)}}{\bigg(K\mu_p + Q_{(s,d,m(\mu_p))}(t)\bigg)}$ (which was formulated in \eqref{eq:genie} in Section \ref{sec:model}) maximizes the network throughput, while attaining low sum costs over flows in the network. Increasing $K$ decreases the cost (by selecting shorter paths), with the price of increasing the queuing delay (by assigning packets to queues with large backlogs). In Section \ref{sec:performance} we show that our novel QUCB's stochastic optimization converges to the optimal solution of the deterministic problem in \eqref{eq:genie} (i.e., with complete knowledge of all path state means) with a logarithmic regret order with time. 

\subsection{Packet travel through the network}
\label{ssec:packets travel}
After leaving the source node $s$ to destination $d$ from the selected queue, packets travel through the network with backpressure policy which directs them to neighbor queues that maximize their differential backlogs. For any node $v$, packets in the $m$th queue (with state $Q_{(v,d,m)}(t)$) need to be delivered to destination node $d$ on a path whose cost is at most $C_m$, and can only be transferred to queues $Q_{(v',d, m')}$ where $v'$ is a neighbor node of $v$, $m'$ is the index of the $m'$th queue in $v'$, and $C_{m'} \le C_m$ (all nodes on the path use the estimated costs which were updated at the same time $t$ for a packet that leaves the source node at time slot $t$). To guarantee these conditions, we define the backpressure parameter as follows. 

At time slot $t$, the backpressure between neighbor queues in nodes $v$ and $v'$ to destination $d$, with queue levels $m$ and $m'$, respectively, is given by:\vspace{0.4cm}
\begin{equation}
    \label{eq:queue_backpressure}
    P_{(v, d, m)}^{(v', d, m')}(t) = 
    \begin{cases}
    Q_{(v, d, m)}(t) - Q_{(v', d, m')}(t), & \vspace{0.4cm}\\
     & \hspace{-3.8cm} C_{m'} \le \max\left(C_m - w_{(v,v')}(t), 0\right),\vspace{0.4cm}\\
    -\infty, & \hspace{-1cm}\mbox{otherwise},
    \end{cases}\vspace{0.4cm}
\end{equation}
and the backpressure of link $(v,v') \in E$, is given by:\vspace{0.4cm}
\begin{equation}
\label{eq:link_backpressure}
    P_{(v,v')}(t) = \max\bigg({{\max\limits_{d,m,m'}{\bigg(P_{(v,d,m)}^{(v',d,m')}(t)\bigg)}, 0}}\bigg).\vspace{0.4cm}
\end{equation}
At time slot $t$, the backpressure parameter $P_{(v,v')}(t)$ of each link $(v,v') \in E$ is evaluated, and link $(v,v^*)$ is selected for transmission, where $v^*=\arg \max_{v'} P_{(v,v')}(t)$. The parameters $d,m,m'$ that solves (\ref{eq:link_backpressure}) for link $(v,v^*)$ state that the next transmitted packet on link $(v,v^*)$ leaves the $m$th queue destined to $d$ at node $v$ and enters the $m'$th queue destined to $d$ at node $v^*$. If the solution of (\ref{eq:link_backpressure}) is zero, then $v$ does not transmit a packet on link $(v, v^*)$ at time slot $t$. Note that in a case of half-duplex transmissions or interference between links, then a certain MAC protocol can be readily applied to manage multi-access transmissions.

Once packets have reached their destination $d$ from source node $s$ through path $p=(s, v_1, v_2, ..., v_I, d)$ at time slot $t$, node $d$ sends an ACK signal back to $s$ through path $p$. In addition to standard operation of acknowledging packet reception, OLSB uses the ACK signal to estimate the path costs, as explained earlier in Subsection \ref{ssec:preprocess}. 

\subsection{Complexity Analysis}
\label{ssec:complexity}

In this section we analyze the computational complexity of the OLSB algorithm. Note that it was shown in \cite{liu2012adaptive}, that by performing a barycentric spanner, the growth in the number of paths is only polynomial (cubic) with the network size instead of an exponential growth of the path complexity in a naive search. As a result, the optimization in (\ref{eq:qucb optimization}) has only polynomial complexity with $|V|$, $O(|V|^3)$, similarly to the path complexity order in \cite{liu2012adaptive} and subsequent studies. Second, in (\ref{eq:queue_backpressure}) and (\ref{eq:link_backpressure}), each node $v$ makes at most $O(NM|\mathcal{D}|)$ computations for backpressure routing, where $N$ is the number of neighbors of $v$, $M$ is the number of queue levels, and $|\mathcal{D}|$ is the number of destinations in the network defined by the network flows, similarly to the backpressure complexity order in \cite{ying2010combining} and subsequent studies.\vspace{0.4cm}

\begin{algorithm}
 \caption{The OLSB Algorithm}\vspace{0.4cm} \label{alg:olsb}
 
 \textit{\textbf{Initialize}}: for every $v \in V$ and every flow destination $d$ do:\vspace{0.4cm}
 
 \textbullet\ Construct a barycentric spanner $\mathcal{B}_{(v,d)}$ out of the path space $\mathcal{P}_{(v,d)}$.\vspace{0.4cm}
 
 \textbullet\ At $t = 0$, transmit one packet through every possible path in $\mathcal{B}_{(v,d)}$, observe the path cost realizations and update $\bar{C}_{p}(0), \forall p \in \mathcal{B}_{(v,d)}$.\vspace{0.4cm}
 
 \textbullet\ Set $T_{p}(0) = 1, \forall p \in \mathcal{B}_{(v,d)}$.\vspace{0.4cm}
 
 For time slot $t \geq 1$, and each flow (say $f_{(s, d)}$) do:\vspace{0.4cm}
 
 \textit{\textbf{Step 1:}} Consider packets arrive at source node $s$ for destination node $d$. Insert the packets to queue $Q_{(s,d,m(\bar{C}_{p^*}(t)))}$, where $p^*$ solves (\ref{eq:qucb optimization}).\vspace{0.4cm}

 \textit{\textbf{Step 2:}} Consider node $v$ in the route is required to transmit packets. Compute the backpressure parameter $P_{(v,v')}(t)$ for every link $(v,v') \in E$ using (\ref{eq:link_backpressure}).\vspace{0.4cm}
 
 \textit{\textbf{Step 3:}} Consider link $(v,v^*)$, where $v^*=\arg \max_{v'} P_{(v,v')}(t)$. If $P_{(v,v^*)} > 0$ and $d, m, m'$ solves (\ref{eq:link_backpressure}) for link $(v,v^*)$, then node $v$ transmits a packet that leaves $Q_{(v,d,m)}$ and enters $Q_{(v^*,d,m')}$ at node $v^*$.\vspace{0.4cm}
 
 \textit{\textbf{Step 4:}} When a packet have reached node $d$ through path $p=(s, v_1, ..., v_I, d)$, update $\bar{C}_{p}(t)$ at node $s$, and $\bar{C}_{p_i}(t)$ of each sub-path $p_i=(v_i,...,v_I,d)$ at node $v_i$. \vspace{0.4cm}
\end{algorithm}

\section{Performance Analysis}
\label{sec:performance}
In this section, we analyze the performance of the OLSB algorithm rigorously theoretically. 

The performance of online learning algorithms are commonly evaluated by the regret, defined as the loss of an algorithm as compared to genie with side information on the system. To evaluate the regret of the OLSB algorithm in this paper, we define a genie with complete knowledge of all path state means, $\mu_p = \sum_{e \in p}{E(w_e(t))}, p \in \mathcal{P}_{(s,d)}$. With this knowledge, genie applies the optimal algorithm, by solving the deterministic optimization problem \eqref{eq:genie}, defined in Section \ref{sec:model}, at each time $t$.

Note that in contrast to existing online learning algorithms for adaptive routing studies, where the optimal solution considers a single and fixed best path in terms of the expected cost (see e.g., \cite{liu2012adaptive, talebi2017stochastic} and references therein), the path selection of the optimal solution in (\ref{eq:genie}) is time-varying due to the queue dynamics. This leads to fundamentally different design and analysis of the learning algorithm. Furthermore, in contrast to weak regret analysis used to simplify the learning design by tracking a static genie, which is restricted to choose the same action over time (see e.g., \cite{auer2002nonstochastic} and subsequent studies in \cite{tekin2012online, liu2012learning, gafni2020learning}), here we aim to minimize a strong regret with respect to genie that takes optimal actions by solving the optimization problem in (\ref{eq:genie}) at each given time, yielding time-varying solutions depending on the queue dynamics. Specifically, we condition on the same queue states for both algorithms, and define the regret $R_n^{OLSB}$ as the loss in performance (the weighted sum of path cost and queue state) attained by OLSB as compared to genie's performance:\vspace{0.4cm}
\begin{equation}
\label{eq:regret_def}
\begin{array}{l}
\displaystyle
R_n^{OLSB} = E\bigg[\sum_{t=1}^{n}{\bigg(KC_{p_t} + Q_{(s,d,m(C_{p_t}(t)))}\bigg)}\bigg] \vspace{0.4cm}\\
\hspace{2cm}
\displaystyle
        -\sum_{t=1}^{n}{\min\limits_{p \in \mathcal{P}_{(s,d)}}{\bigg(K\mu_p + Q_{(s,d,m(\mu_p))}\bigg)}},
        \vspace{0.4cm}
\end{array}
\end{equation}
where $p_t$ is the actual path chosen by OLSB at time slot $t$, and $C_{p_t} = \sum_{e \in p}{w_e(t)}$ is the actual path cost incurred through path $p_t$ at time slot $t$. 

In the following theorem we establish the upper bound on the regret $R_n^{OLSB}$ for each flow and for all $n$ and show that it has a logarithmic order with time.\vspace{0.4cm} 

\begin{theorem}
\label{th:regret}
The regret $R_n^{OLSB}$ is upper bounded by:\vspace{0.4cm}
\begin{equation}
\displaystyle    \bigg[8\sum_{\substack{i=1,...,L\\i: \Delta_i^{min}\neq0}}{\frac{\Psi_i\ln{n}}{(\Delta_i^{min})^2}}\bigg] + (L - 1)(1 + \frac{\pi^2}{3})\sum_{i=1}^{L}{\Psi_i},\vspace{0.1cm}
\end{equation}\vspace{0.4cm}
\end{theorem}

where\vspace{0.4cm}
\begin{equation}
\Psi_i \triangleq (K\mu_i + \eta_{m(\mu_i)}) - (K\min\limits_{\substack{j=1,...,L\\j \neq i}}{\mu_j} + \min\limits_{\substack{j=1,...,L\\j \neq i}}{\eta_{m(\mu_j)}}),\vspace{0.4cm}
\end{equation}

\begin{equation}
\Delta_i^{min} \triangleq \min\limits_{\substack{j=1,...,L\\j\neq i}}{\bigg((K\mu_j + \eta_{m(\mu_j)}) - (K\mu_i + \eta_{m(\mu_i)})\bigg)},\vspace{0.4cm} 
\end{equation}
$\ln(\cdot)$ is the natural logarithm, $L$ is the number of barycentric spanner paths of the flows, and $\eta_{m(\mu_i)}$ is the mean value of queue $m(\mu_i)$ at the source node.\vspace{0.4cm}

The proof is given in the Appendix.

\section{Simulation Results}
\label{sec:experiemnts}

\begin{center}
    \begin{figure}[ht]
        \centering
        \includegraphics[width=8cm]{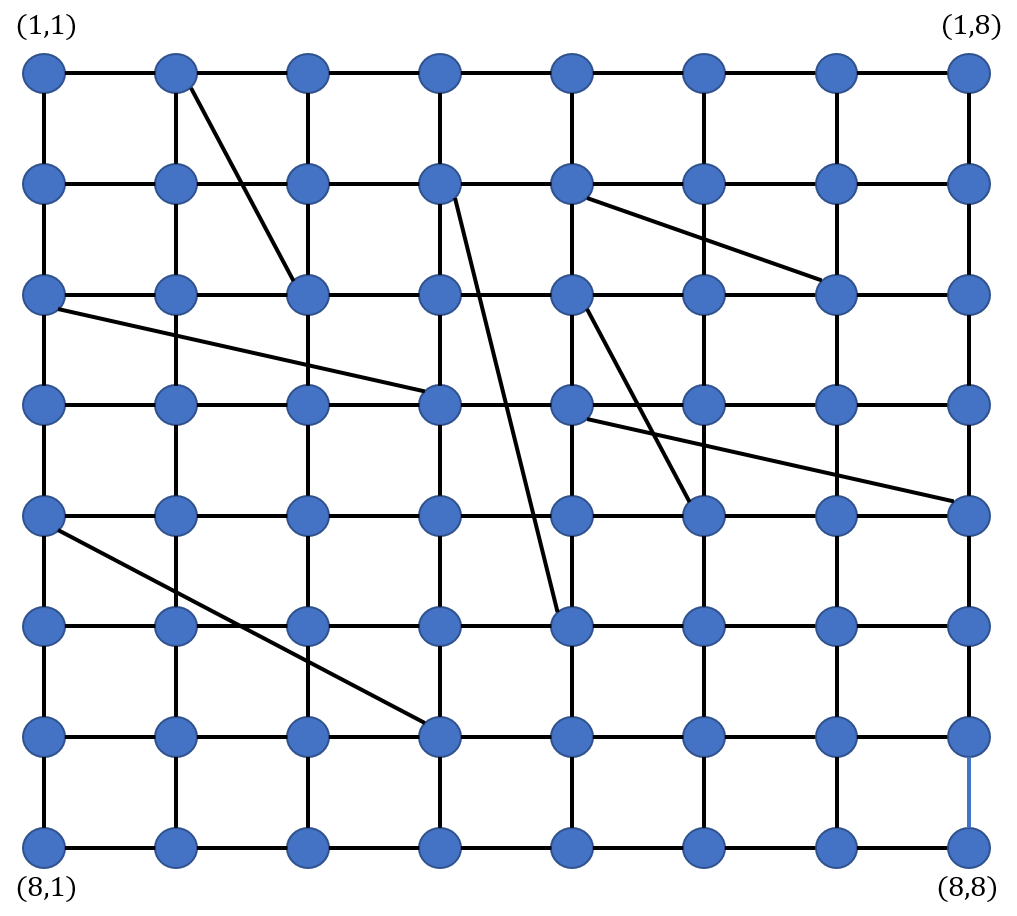}
        \captionsetup{font=footnotesize}
        \caption{An illustration of the network used in the simulations.}
        \label{fig:network}
    \end{figure}
\end{center}

In this section, we present simulation results to validate the theoretical findings, and demonstrate the efficiency of the OLSB algorithm. We simulated a similar directed network as in \cite{ying2010combining} with $64$ nodes and $119$ links. An illustration is shown in Fig. \ref{fig:network}. The additional links were inserted to model the case of different hopping transmissions (as in 5G mesh networks).
We simulated nine flows in the networks as shown in Table \ref{table:flows}, such that two flows originate in the same source node, two flows are targeted to the same destination node and five random flows. The packet arrivals of all flows follow a Poisson process with rate $\lambda$. At the beginning of the simulations, all queues are empty. 

\subsection{Evaluating the Convergence of OLSB to the Optimal Strategy \cite{ying2010combining}}

In this simulations we demonstrate the learning efficiency of OLSB as compared to the optimal solution by genie that has complete knowledge of the path state means\cite{ying2010combining}. 

\begin{figure}[ht]
  \centering
  \sbox\figureone{\includegraphics[height=7cm]{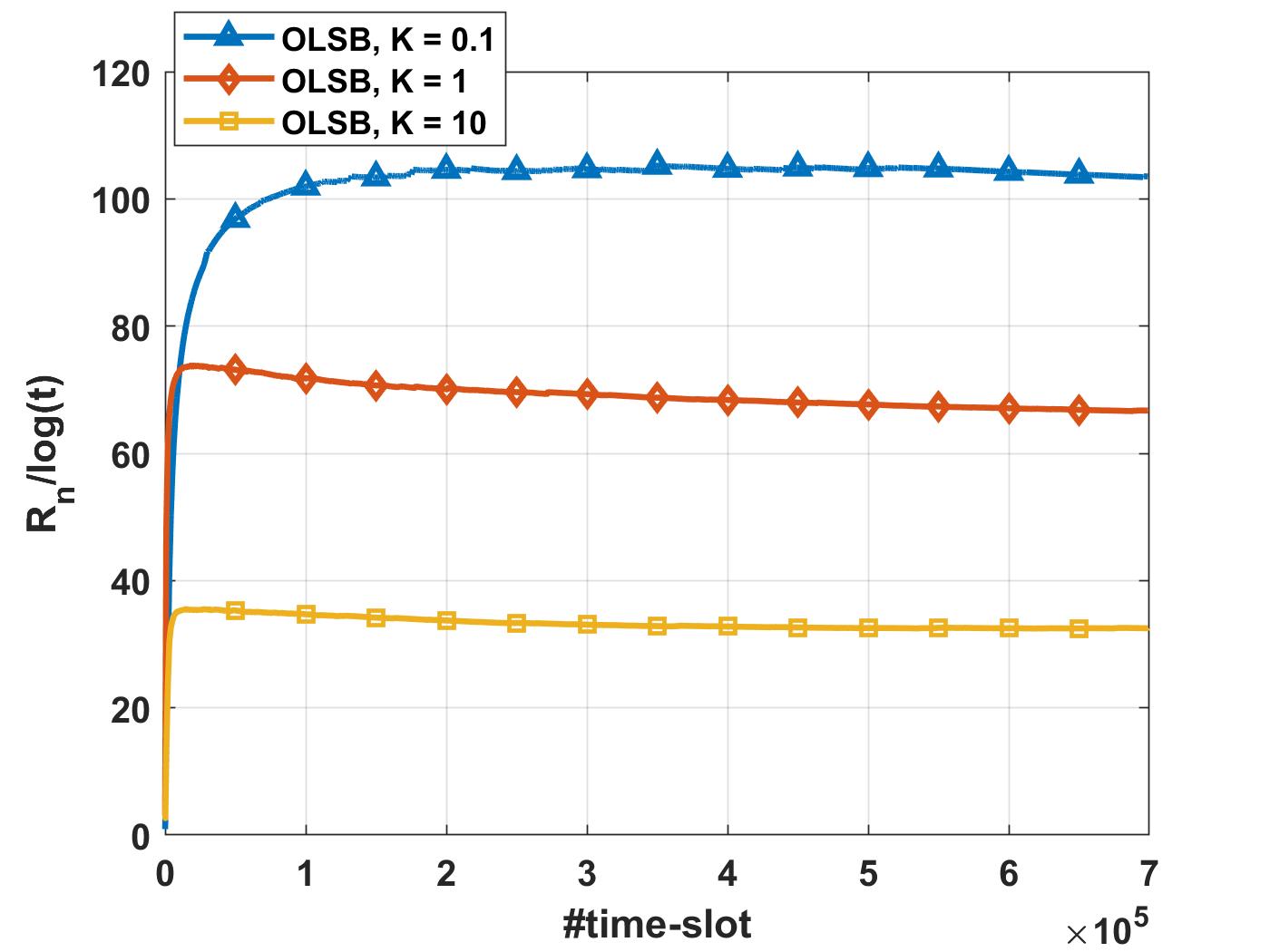}}%
  \usebox\figureone
  \par
  \begin{minipage}{\wd\figureone}
  \captionsetup{font=footnotesize}
  \caption{The empirical regret (normalized by $\log{t}$) obtained by OLSB algorithm as compared to genie's performance, for $K = [0.1, 1, 10]$ and $\lambda = 1$.}
  \label{fig:regret_compare_K}
  \end{minipage}
\end{figure}

\vspace*{-\baselineskip}
\vspace*{-\baselineskip}
\begin{center}
\begin{table}[ht]
    \large
    \centering
    \begin{tabular}{||c c c||}
         \hline
         Flow & Source & Destination \\ [0.5ex] 
         \hline\hline
         1 & (1, 2) & (4, 4) \\ 
         \hline
         2 & (1, 2) & (8, 4) \\ 
         \hline
         3 & (2, 2) & (3, 7) \\
         \hline
         4 & (2, 6) & (8, 8) \\
         \hline
         5 & (3, 3) & (8, 6) \\
         \hline
         6 & (3, 4) & (5, 8) \\
         \hline
         7 & (4, 1) & (6, 8) \\
         \hline
         8 & (5, 3) & (7, 8) \\
         \hline
         9 & (5, 4) & (8, 8) \\
         \hline
    \end{tabular}
    \captionsetup{font=footnotesize}
    \captionof{table}{The locations of flows used in the simulations. Flows $1$ and $2$ both originate at node $(1, 2)$; flows $4$ and $9$ both destined to reach node $(8, 8)$.}
    \label{table:flows}
\end{table}
\end{center}
We start by validating the theoretical analysis of the regret, which measures the convergence speed of OLSB to the optimal strategy. For this, we computed the regret empirically according to (\ref{eq:regret_def}) and normalized it by $\log(t)$ (i.e., converging to a constant value validates the logarithmic order of the regret with time). In Fig. \ref{fig:regret_compare_K}, we show the influence of the selection of the $K$ parameter on the regret curve. We note that since the coefficient of the logarithm in the regret expression is inversely proportional to the value of $K$, lower values of $K$ result in a longer convergence time. This means that it is easier to learn strategies that assign high priority for transmissions over short paths. This observation is intuitively satisfying, as the algorithm is required to learn smaller subsets of path selections. It can be seen clearly that we obtained a logarithmic regret order with time for each selection of $K$, which supports the theoretical results.

\subsection{Evaluating the Latency and Network Congestion under OLSB}

Next, we evaluate the latency and network congestion achieved by the OLSB algorithm. To this, we present the average end-to-end delay of all successful transmissions, side-by-side with the average per-node queue lengths, for different selections of the $K$ parameter, and for low, moderate and high loads. We set the time slot duration to $20{\mu}s$. We compare the results with the following routing methods: The backpressure routing algorithm, that routes data in directions that maximize the differential queue backlog between nodes to reduce the congestion \cite{moeller2010routing}, the Adaptive Shortest Path Routing (ASPR) algorithm, that uses adaptive strategies to learn the shortest path routing \cite{liu2012adaptive}, and the recently suggested reinforcement learning routing method that uses multi-armed bandit framework based on UCB1 for path learning and packet transmissions (RL-UCB1) \cite{tabei2021multi}.

\begin{figure}[ht]
 \centering
 \vspace*{5pt}%
   \begin{subfigure}{0.5\textwidth}
        \centering
        \includegraphics[height=6cm]{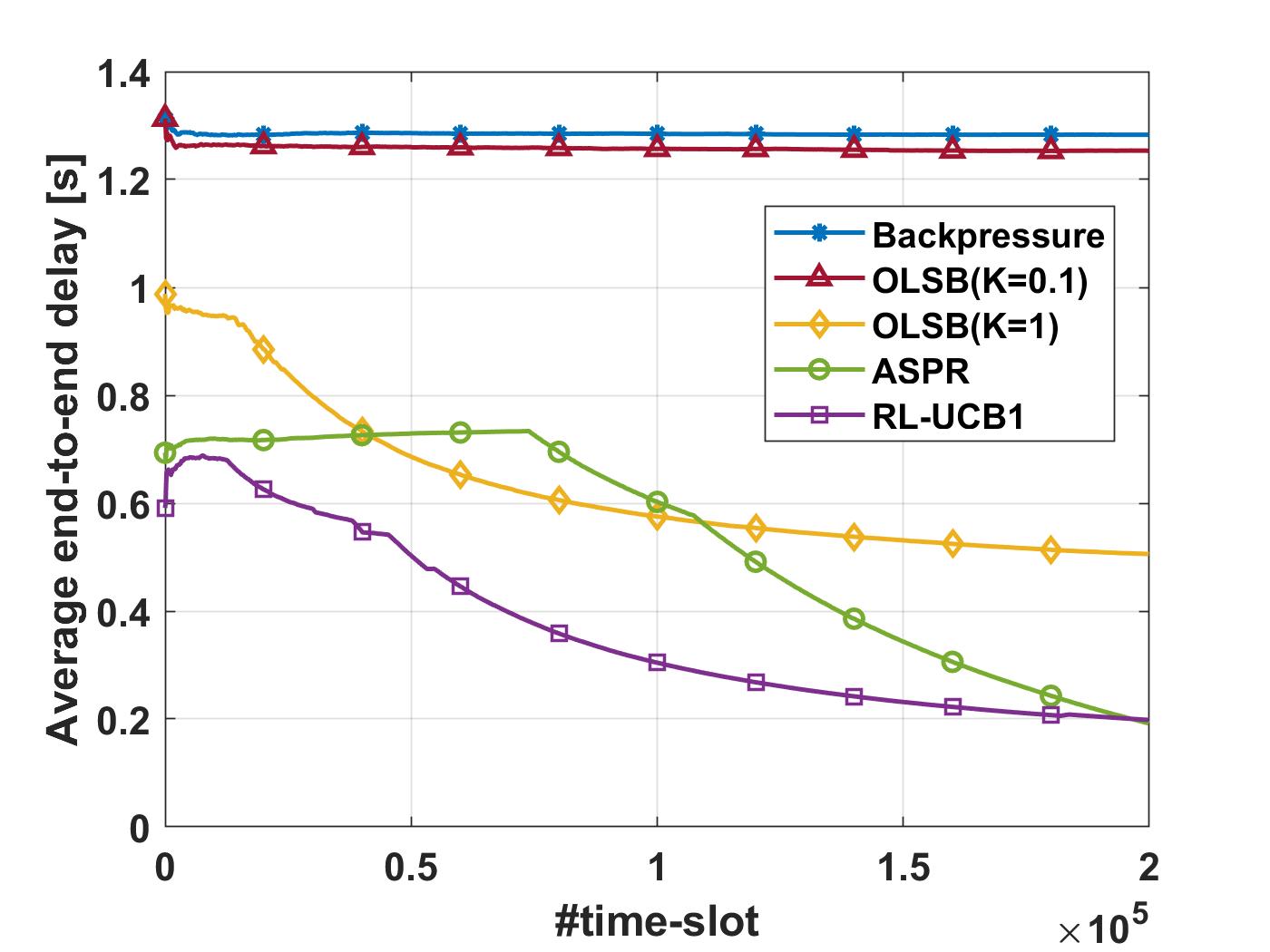}%
        \captionsetup{font=footnotesize}
        \caption{Average end-to-end delay.}
        \label{subfig:delay_low_load}
   \end{subfigure} 
   
   \vspace*{8pt}%
   
   \begin{subfigure}{0.5\textwidth}
        \centering
        \includegraphics[height=6cm]{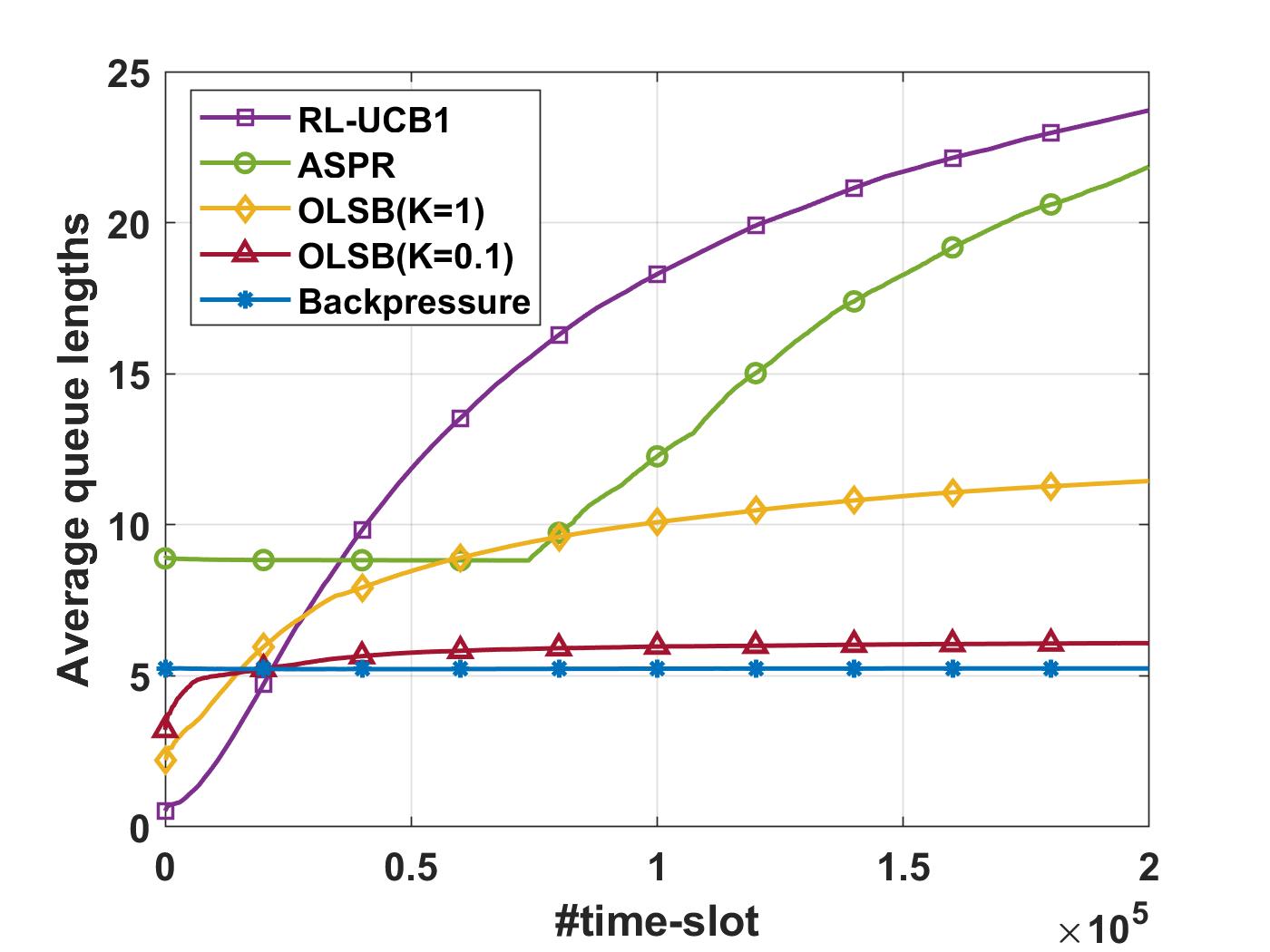}%
        \captionsetup{font=footnotesize}
        \caption{Average per-node queue lengths.}
        \label{subfig:queue_low_load}
   \end{subfigure}     
\captionsetup{font=footnotesize}
\caption{The average end-to-end delay and the average per-node queue length in a lightly-loaded network ($\lambda = 0.6$).}
\label{fig:low_load}
\end{figure}

In Fig. \ref{fig:low_load}, we present simulation results of a lightly-loaded network ($\lambda = 0.6$). It is shown that setting larger $K$ values in OLSB leads to better performance in terms of average end-to-end delay but results in higher queue loads. This is because large $K$ values leads to more frequent selections of short paths by increasing this priority in the objective function. However, we note that this is an acceptable behaviour of low arrival rates, since the exploration of longer paths is not necessary for load balancing. Furthermore, as discussed above, the backpressure algorithm performs poorly under light loads because of extensive and unnecessary exploration of paths for network stability. Therefore, while backpressure routing remains stable over time, it does not exploit better paths, in terms of the total cost, as the OLSB algorithm. The ASPR and RL-UCB1 tends to be unstable over time, as they fail to balance the congestion in the network.

\begin{figure}[ht]
 \centering
 \vspace*{5pt}%
   \begin{subfigure}{0.5\textwidth}
        \centering
        \includegraphics[height=6cm]{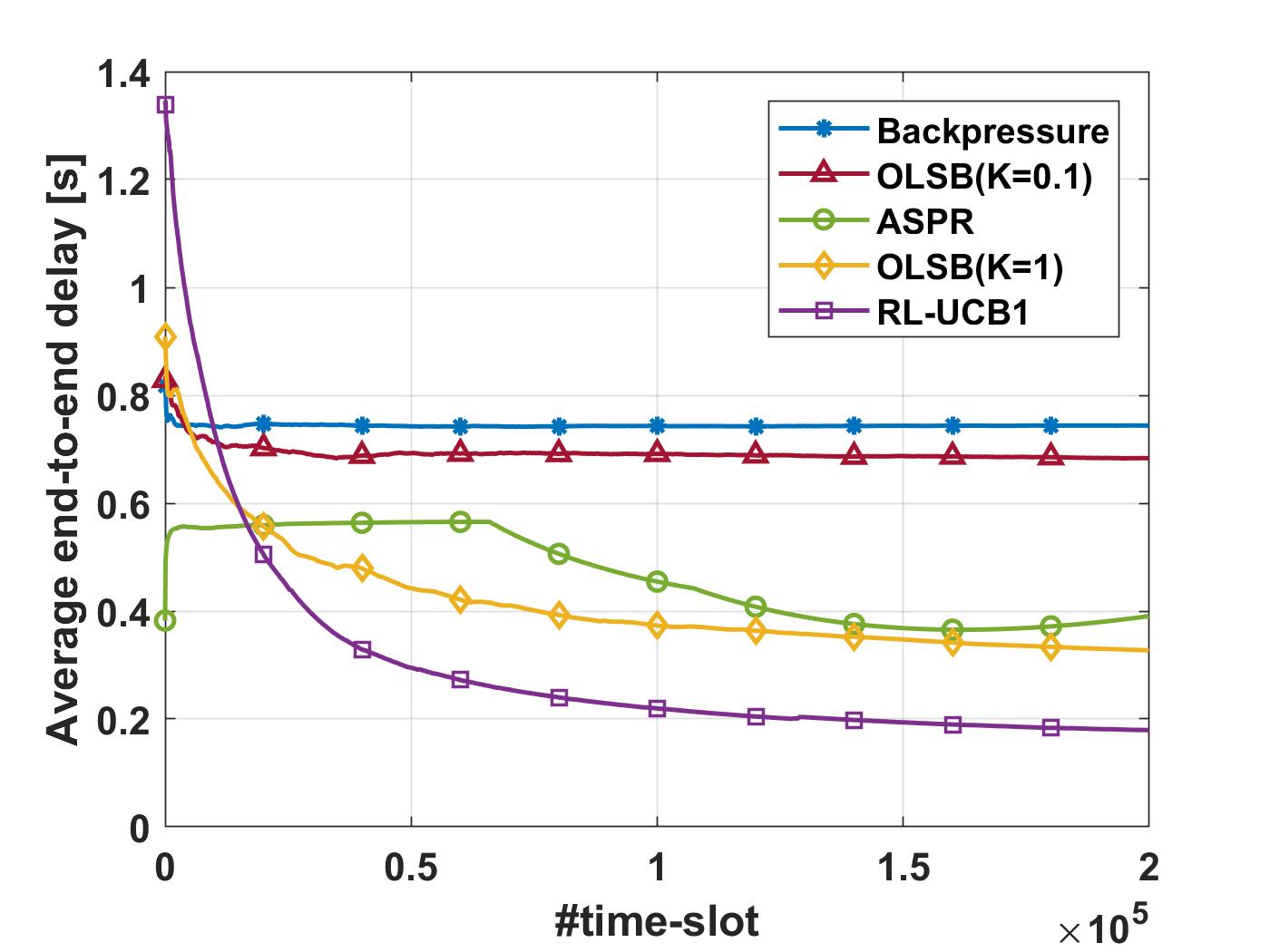}%
        \captionsetup{font=footnotesize}
        \caption{Average end-to-end delay.}
        \label{subfig:delay_moderate_load}
   \end{subfigure} 
   
   \vspace*{8pt}%
   
   \begin{subfigure}{0.5\textwidth}
        \centering
        \includegraphics[height=6cm]{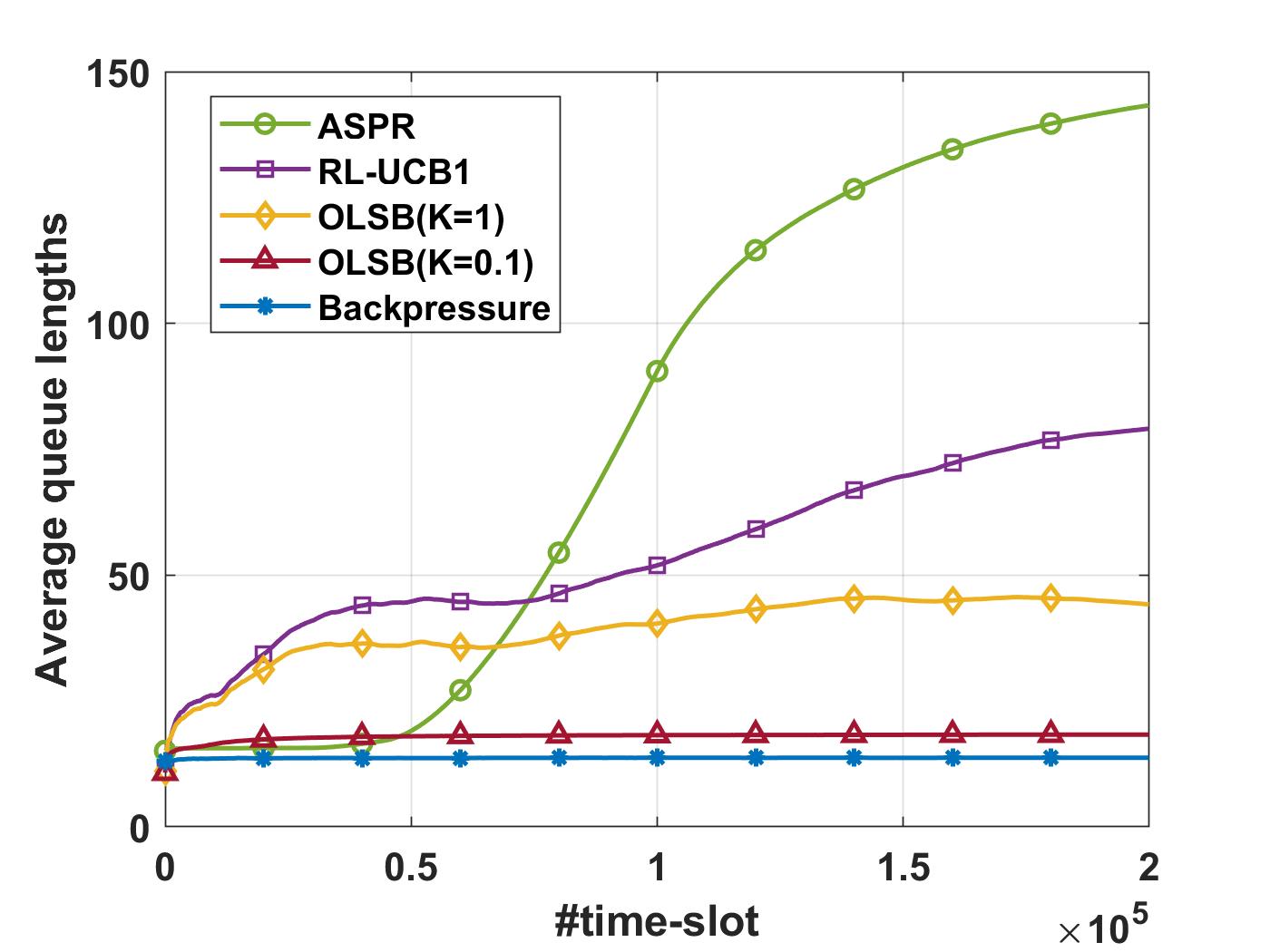}%
        \captionsetup{font=footnotesize}
        \caption{Average per-node queue lengths.}
        \label{subfig:queue_moderate_load}
   \end{subfigure}     
\captionsetup{font=footnotesize}
\caption{The average end-to-end delay and the average per-node queue length in a moderately-loaded network ($\lambda = 1$).}
\label{fig:moderate_load}
\end{figure}

In fig. \ref{fig:moderate_load}, we present simulation results of a moderately-loaded network ($\lambda = 1$). We obtained a similar behaviour of OLSB as in the lightly-loaded network, as it still performs well. It can be seen that the improvement of the OLSB algorithm over the backpressure algorithm increases. The RL-UCB1 performs well in this scenario as well, although its stability is limited. The ASPR tends to be unstable over time.

\begin{figure}[ht]
 \centering
 \vspace*{5pt}%
   \begin{subfigure}{0.5\textwidth}
        \centering
        \includegraphics[height=6cm]{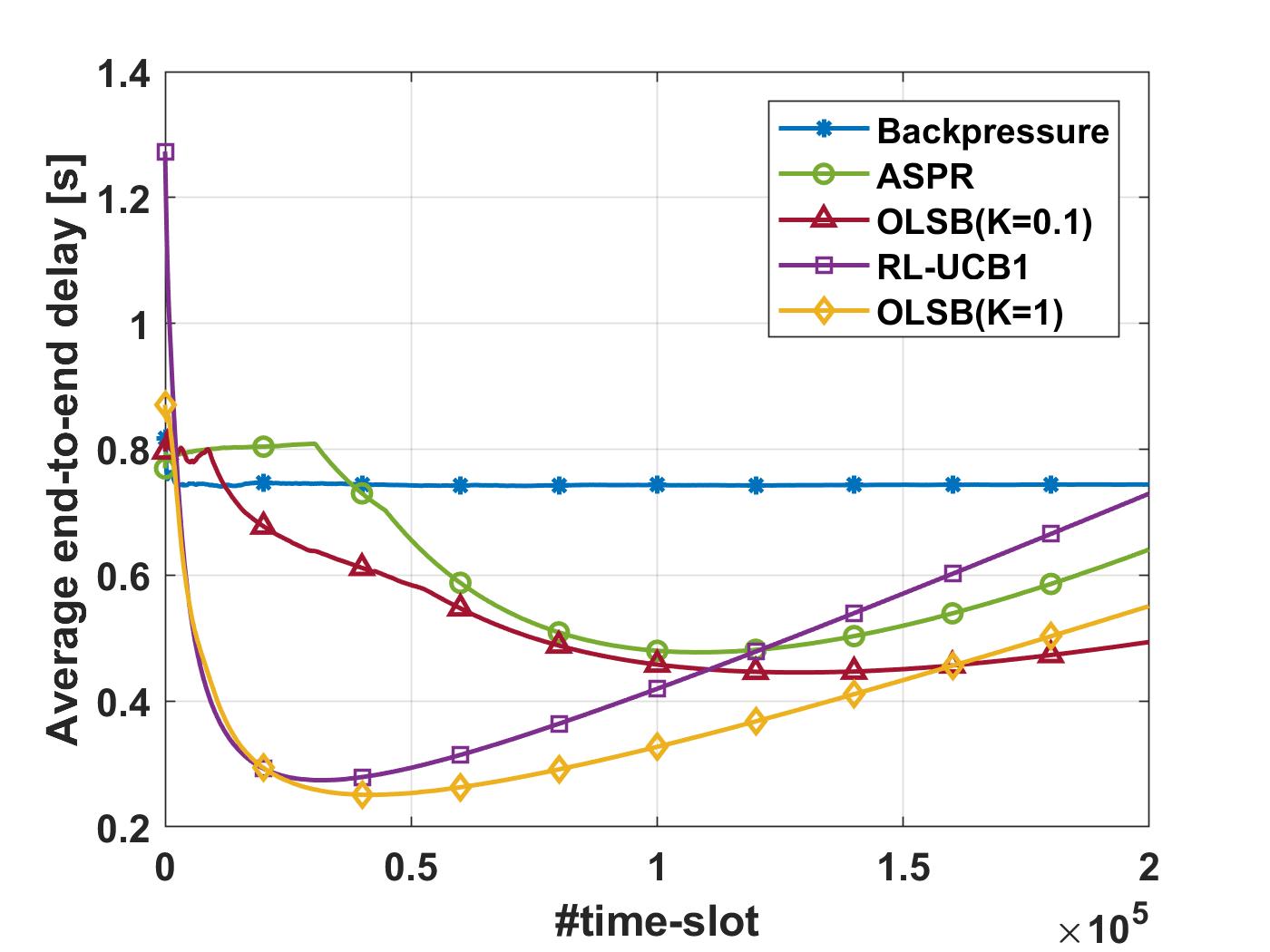}%
        \captionsetup{font=footnotesize}
        \caption{Average end-to-end delay.}
        \label{subfig:delay_high_load}
   \end{subfigure} 
   
   \vspace*{8pt}%
   
   \begin{subfigure}{0.5\textwidth}
        \centering
        \includegraphics[height=6cm]{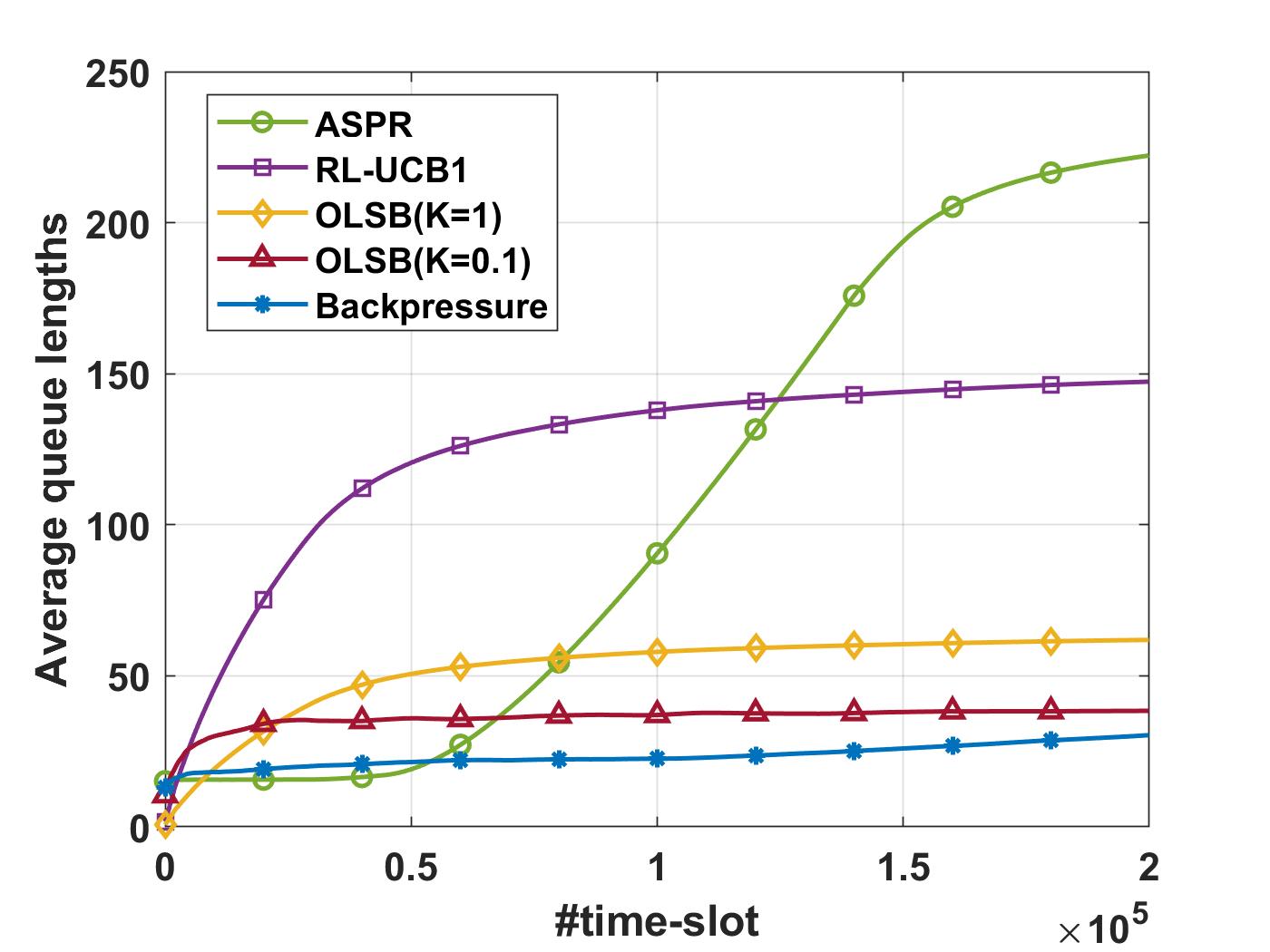}%
        \captionsetup{font=footnotesize}
        \caption{Average per-node queue lengths.}
        \label{subfig:queue_high_load}
   \end{subfigure}     
\captionsetup{font=footnotesize}
\caption{The average end-to-end delay and the average per-node queue lengths in a highly-loaded network ($\lambda = 1.5$).}
\label{fig:high_load}
\end{figure}

Finally, we simulated a highly-loaded network ($\lambda = 1.5$). The results are presented in Fig. \ref{fig:high_load}. It can be seen that both RL-UCB1 and OLSB (with $K=1$) learn the path cost quickly. However, it can be inferred that the shortest-path queues are filled quickly as well and the delay grows with time. This is an undesired behaviour since sub-optimal queues are rarely used. In this case, it can be seen that OLSB with $K=0.1$ shows strong performance in terms of end-to-end delay as well as queue stability. This obtained by reducing the priority of using short paths when decreasing $K$ in the OLSB optimization. This is intuitively satisfying, as increasing the priority of backpressured transmissions together with efficient path exploration and exploitation mechanism of the OLSB optimization is desired in high loads. Finally, the pure backpressure algorithm shows balanced behaviour, as expected when all queues are utilized equally.
Moreover, it can be seen that OLSB with $K=0.1$ achieves low congestion level compared to the other algorithms. As expected, both RL-UCB1 and ASPR perform poorly under high loads in terms of average queue length since they highly prioritize transmissions through short paths rather than transmissions that achieve efficient queue balancing. Furthermore, it can be seen that OLSB outperforms the RL-UCB1 and  backpressure algorithms. This is because RL-UCB1 learns a fixed set of paths across time, while OLSB balances between the minimal cost and the time-varying queue states. Also, the backpressure algorithm results in sending packets in long paths, which reduces the performance in terms of end-to-end delay.

\section{Conclusion}
\label{sec:conclusion}
We considered the problem of adaptive routing under unknown path states. We developed a novel Online Learning for Shortest path and Backpressure (OLSB) algorithm to maximize the network throughput (i.e., support the capacity region by using backpressured paths) while attaining low sum costs over flows in the network (by using short paths). We have analyzed OLSB theoretically and showed that it attained a logarithmic regret order as compared to a genie that has complete knowledge of the path state means. We presented simulation results that support the theoretical findings, and demonstrate strong performance of the OLSB algorithm. Specifically, OLSB demonstrated strong and robust performance in all simulations, while other existing methods failed to present robust performance. Furthermore, OLSB has the ability of optimizing the performance depending on the network load by adjusting a simple tuning parameter that controls the balancing between using short paths and reducing the congestion level, which makes it simple for implementation in practical networks.  

\section{Appendix}
In this appendix, we provide the proof of Theorem \ref{th:regret}.

Throughout the proof we denote the source node and destination node of the flow by $s$, $d$, respectively. The selected path by OLSB at time $t$ is denoted by $p_t$.
and let\vspace{0.4cm} 
\begin{center}
$g_t\triangleq \argmin\limits_{p \in \mathcal{P}_{(s,d)}}{\{\mu_p + Q_{(s,d,m(\mu_p))}(t)\}}$\vspace{0.4cm}
\end{center}
be the optimal path which is selected by genie at time step $t$.

\noindent
The cumulative regret after $n$ plays is given by:\vspace{0.4cm}
\begin{equation}
    \label{eq:regret}
\begin{array}{l}
        \displaystyle R_n= E\Bigg[\sum_{t=1}^{n}{\bigg(K{C}_{p_t}(t) + Q_{(s,d,m({C}_{p_t}(t)))}(t)\bigg)} \vspace{0.4cm}\\\hspace{1.5cm}
        \displaystyle
        - \sum_{t=1}^{n}{K{C}_{g_t}(t) + Q_{(s,d,m({C}_{g_t}(t)))}(t)\bigg)}\bigg].\vspace{0.4cm}
\end{array}    
\end{equation}

\noindent
We can rewrite the first term on the RHS of (\ref{eq:regret}) by summing the balanced cost over paths:

\begin{equation}
    \label{eq:first_term}
\begin{array}{l}
        \displaystyle E\Bigg[\sum_{t=1}^{n}{\bigg(K{C}_{p_t}(t) + Q_{(s,d,m({C}_{p_t}(t)))}(t)\bigg)}\bigg] \vspace{0.4cm}\\\hspace{1cm}
\displaystyle=\sum_{p \in \mathcal{P}_{(s,d)}}{(K\mu_p + \eta_{m(\mu_p)})E[T_p(n)]}.\vspace{0.4cm}
\end{array}
\end{equation}

Next, we can bound the second term on the RHS of (\ref{eq:regret}) by using the linearity of expectation and summing the minimum over $K\mu_p$ plus the minimum over $\eta_{m(\mu_p)}$ at each time $t$:

\begin{equation}
    \label{eq:second_term}
\begin{array}{l}
\displaystyle        E\Bigg[\sum_{t=1}^{n}{K{C}_{g_t}(t) + Q_{(s,d,m({C}_{g_t}(t)))}(t)\bigg)}\bigg] \geq \vspace{0.4cm}\\\hspace{1cm}
\displaystyle 
        \bigg(\min\limits_{p \in \mathcal{P}_{(s,d)}}{K\mu_p} + \min\limits_{p \in \mathcal{P}_{(s,d)}}{\eta_{m(\mu_p)}}\bigg)n.\vspace{0.4cm}
\end{array}    
\end{equation}

By substituting (\ref{eq:first_term}) and (\ref{eq:second_term}) in (\ref{eq:regret}), we can upper bound the cumulative regret by:\vspace{0.4cm}

\begin{equation}
\label{eq:reg_bound}
\begin{array}{l}
        \displaystyle 
        R_n \leq \sum_{p \in \mathcal{P}_{(s,d)}}{(K\mu_p + \eta_{m(\mu_p)})E[T_p(n)]} - \vspace{0.4cm}\\\hspace{1cm}
        \displaystyle 
         - \bigg(\min\limits_{p \in \mathcal{P}_{(s,d)}}{K\mu_p} + \min\limits_{p \in \mathcal{P}_{(s,d)}}{\eta_{m(\mu_p)}}\bigg)n.\vspace{0.4cm}
\end{array}
\end{equation}

We next upper bound the expected value of the number of times that path $p$ was selected for transmission. Let 
\begin{equation}
\displaystyle
c_{t,s}\triangleq\sqrt{\frac{2\ln{t}}{s}}.
\end{equation}
Then,\vspace{0.4cm} 
\begin{equation*}
\begin{array}{l}
\displaystyle    
T_p(n) =_{\textbf{(a)}} 1+\sum_{t=L+1}^{n}{\bigg\{p_t = p\bigg\}}\vspace{0.4cm}\\\hspace{1cm}
\displaystyle
     \leq_{\textbf{(b)}} l+\sum_{t=L+1}^{n}{\bigg\{p_t = p, T_p(t-1)\geq l\bigg\}}\vspace{0.4cm}
\end{array}
\end{equation*}
\begin{equation*}
\begin{array}{l}
\displaystyle    
     \\\hspace{1cm}
\displaystyle
     \leq_{\textbf{(c)}} l+\sum_{t=L+1}^{n}\bigg\{K\underbrace{\bar{C}_{g_{t-1}}(T_{g_{t-1}}(t-1))}_{\bar{\triangleq C}_g}\vspace{0.4cm}\\
     \hspace{2cm} + Q_{(s,d,m(\bar{C}_g))}(t-1) - c_{t-1,T_{g_{t-1}}(t-1)}\vspace{0.4cm}\\\hspace{2cm}
     \geq K\underbrace{\bar{C}_p(T_p(t-1))}_{\triangleq\bar{C}_p} + Q_{(s,d,m(\bar{C}_p))}(t-1)\vspace{0.4cm}\\\hspace{3cm}
     - c_{t-1,T_{p}(t-1)}, T_p(t-1)\geq l\bigg\}\vspace{0.4cm}\\
\end{array}
\end{equation*}
\begin{equation*}
\begin{array}{l}
\displaystyle    
    \leq_{\textbf{(d)}} l + \sum_{t=L+1}^{n}\bigg\{\max\limits_{0<s_g<t}\min\limits_{\substack{r \in \mathcal{P}_{(s,d)}\\r \neq p}}K\bar{C}_r(s_g)\vspace{0.4cm}\\
\displaystyle    
    \hspace{2.5cm} + Q_{(s,d,m(\bar{C}_r(s_g)))}(s_g) - c_{t-1,s_g}\vspace{0.4cm}\\ 
\displaystyle    
    \hspace{2cm} \geq \min\limits_{l\leq s_p\leq t}K\bar{C}_{p}(s_p) + Q_{(s,d,m(\bar{C}_{p}(s_p)))}(s_p) \vspace{0.4cm}\\
\displaystyle   
\hspace{7cm} - c_{t-1,s_p}\bigg\} 
\vspace{0.4cm}
\end{array}
\end{equation*}
\begin{equation}
\label{eq:Tp_bound}
\begin{array}{l}
\displaystyle
     \leq_{\textbf{(e)}} l + \sum_{t=L + 1}^{n}\sum_{s_g=1}^{t-1}\sum_{s_p=l}^{t-1}\sum_{\substack{r \in \mathcal{P}_{(s,d)}\\r \neq p}}\vspace{0.4cm}\\
     \displaystyle
     \hspace{2cm} \bigg\{K\bar{C}_r(s_g)+ Q_{(s,d,m(\bar{C}_r(s_g)))}(s_g) - c_{t-1,s_g}\vspace{0.4cm}\\
     \displaystyle
    \hspace{3cm} \geq K\bar{C}_{p}(s_p) + Q_{(s,d,m(\bar{C}_{p}(s_p)))}(s_p) \vspace{0.4cm}\\  
    \displaystyle
     \hspace{7cm} - c_{t-1,s_p}\bigg\},\vspace{0.4cm}
\end{array}    
\end{equation}

where $\bigg\{E\bigg\}$ is the indicator function, which equals $1$ when event $E$ is true, and equals $0$ otherwise. Below, we explain each bounding step of $T_p(n)$ in \eqref{eq:Tp_bound}:\vspace{0.4cm}

\begin{enumerate}[label=(\alph*)]
\item Step (a) follows since the number of times that path $p$ was selected for transmission up to time $n$ is given by the sum of one (due to the first initial path selection) plus the number of time-slots in which path $p$ was selected by the algorithm, i.e. $p^*(t) = p$. \vspace{0.4cm} 

\item Step (b) follows since we take $l - 1$ occurrences out of the sum and condition the sum to count path $p$ selection only after it was selected $l$ times.\vspace{0.4cm}

\item Step (c) follows since the event $p_t = p$ occurs when $p$ solves the QUCB rule in OLSB:\vspace{0.4cm} 
\begin{center}
    $p = \argmin\limits_{r \in \mathcal{P}_{(s,d)}}{K\bar{C}_r(t) + Q_{(s,d,m(\bar{C}_r(t)))}(t) + c_{t,T_r(n)}}$.\vspace{0.4cm}
\end{center}
Also, note that by the definition of minimization the solution is smaller or equal than the value of the function when the argument is path $g_t$ which was selected by genie, which yields Step (c).
\vspace{0.4cm}

\item Step (d) further upper bounds the expression since if the value of path $p$ is smaller than the value of path $g_{t-1}$ then its minimal value from time $l$ to the current time $t$ is smaller than the maximum over all minimal values by other path selections up to time $t$. When the condition holds, we get one triplet of $(r, s_g, s_p)$ that we count as path $p$ selection.\vspace{0.4cm}

\item Step (e) follows since we count every $(r, s_g, s_p)$ triplet that meets the condition.\vspace{0.4cm}
\end{enumerate}
Next, note that for condition 
\begin{equation}
\label{eq:p_selection}
\begin{array}{l}
K\bar{C}_r(s_g) + Q_{(s,d,m(\bar{C}_r(s_g)))}(s_g) - c_{t,s_g} \vspace{0.4cm}\\\hspace{1cm}
\geq K\bar{C}_{p}(s_p) + Q_{(s,d,m(\bar{C}_{p}(s_p)))}(s_p) - c_{t,s_p}
\end{array}
\end{equation}
to hold, then for each $r \neq p$ at least one of the following inequalities must hold:\vspace{0.4cm}

\noindent
\emph{Inequality 1:}
    \begin{equation}
    \label{eq:cond_1}
    \begin{array}{l}
\displaystyle         
K\bar{C}_r(s_g) + Q_{(s,d,m(\bar{C}_r(s_g)))}(s_g)  \vspace{0.4cm}\\
\hspace{2cm}
\displaystyle
         \geq K\mu_r + Q_{(s,d,m(\bar{C}_r(s_g)))}(s_g) + c_{t,s_g}.
    \end{array}
    \end{equation}

\noindent
\emph{Inequality 2:}
    \begin{equation}
    \label{eq:cond_2}
    \begin{array}{l}
    \displaystyle
         K\bar{C}_{p}(s_p) + Q_{(s,d,m(\bar{C}_{p}(s_p)))}(s_p)  
\vspace{0.4cm}\\
\hspace{2cm}
\displaystyle
\leq K\mu_p + Q_{(s,d,m(\bar{C}_{p}(s_p)))}(s_p) - c_{t,s_p}.
    \end{array}
    \end{equation}

\noindent
\emph{Inequality 3:}
    \begin{equation}
    \label{eq:cond_3}
    \begin{array}{l}
         K\mu_r + Q_{(s,d,m(\bar{C}_r(s_g)))}(s_g)  
\vspace{0.4cm}\\
\hspace{2cm}
\displaystyle
         > K\mu_p + Q_{(s,d,m(\bar{C}_{p}(s_p)))}(s_p) - 2c_{t,s_p}.
    \end{array}
    \end{equation}
Therefore, we get $L - 1$ sets of these three inequalities.

We prove by contradiction that by assuming that if for all $r \in \mathcal{P}_{(s,d)}, r \neq p$ all inequalities are false, then:\vspace{0.4cm}
\begin{equation}
\label{eq:k_mu_r_plus_Q}
    \begin{array}{l}
        K\mu_r + Q_{(s,d,m(\bar{C}_r(s_g)))}(s_g) \vspace{0.4cm}\\ 
\displaystyle
\hspace{1cm}
        >_{\textbf{(a)}}  K\bar{C}_r(s_g) + Q_{(s,d,m(\bar{C}_r(s_g)))}(s_g) - c_{t,s_g} 
\vspace{0.4cm}\\ 
\displaystyle
\hspace{1cm}
        \geq_{\textbf{(b)}} K\bar{C}_{p}(s_p) + Q_{(s,d,m(\bar{C}_{p}(s_p)))}(s_p) - c_{t,s_p} 
\vspace{0.4cm}\\ 
\displaystyle
\hspace{1cm}
>_{\textbf{(c)}} K\mu_p + Q_{(s,d,m(\bar{C}_{p}(s_p)))}(s_p) - 2c_{t,s_p}.\vspace{0.4cm}
    \end{array}
\end{equation}
Below, we explain each bounding step in \eqref{eq:k_mu_r_plus_Q}:\vspace{0.4cm}
\begin{enumerate}[label=(\alph*)]
\item Step (a) follows by assuming that inequality (\ref{eq:cond_1}) is false.\vspace{0.4cm} 

\item Step (b) follows by inequality (\ref{eq:p_selection}).\vspace{0.4cm}

\item Step (c) follows by assuming that inequality (\ref{eq:cond_2}) is false.\vspace{0.4cm}
\end{enumerate}

\noindent
Therefore, we get:
\begin{equation}
\begin{array}{l}
K\mu_r + Q_{(s,d,m(\bar{C}_r(s_g)))}(s_g) 
\vspace{0.4cm}\\
\hspace{1cm}
\displaystyle
< K\mu_p + Q_{(s,d,m(\bar{C}_{p}(s_p)))}(s_p) - 2c_{t,s_p}
\end{array}
\end{equation}
which meets inequality (\ref{eq:cond_3}), which is in contradiction to the assumption that all three inequalities are false.

Next, we apply the Chernoff-Hoeffding bound on inequalities (\ref{eq:cond_1}) and (\ref{eq:cond_2}), and get: 

\begin{equation}
\begin{array}{l}
Pr(K\bar{C}_r(s_g) - K\mu_r\geq c_{t,s_g})  
\vspace{0.4cm}\\\hspace{1cm}
\displaystyle
\leq e^{-2s_g c_{t,s_g}^2} = e^{-2s_g\frac{2\ln{t}}{s_g}} = t^{-4},
\end{array}
\end{equation}
and
\begin{equation}
\begin{array}{l}
Pr(K\mu_p - K\bar{C}_{p}(s_p) \geq c_{t,s_p}) 
\vspace{0.4cm}\\\hspace{1cm}
\displaystyle
\leq e^{-2s_p c_{t,s_p}^2} = e^{-2s_p\frac{2\ln{t}}{s_p}} = t^{-4}.
\end{array}
\end{equation}
Also, it suffices to choose inequality (\ref{eq:cond_3}) to be false:

\begin{equation}
\begin{array}{l}
K\mu_r + Q_{(s,d,m(\bar{C}_r(s_g)))}(s_g) 
\vspace{0.4cm}\\\hspace{1cm}
\displaystyle
\leq K\mu_p + Q_{(s,d,m(\bar{C}_{p}(s_p)))}(s_p) - 2c_{t,s_p}.
\end{array}
\end{equation}
We take expectation and get:
\begin{equation}
\begin{array}{l}
K\mu_r + \eta_{m(\mu_r)} \leq K\mu_p + \eta_{m(\mu_p)} - 2c_{t,s_p},
\end{array}
\end{equation}
and by arranging terms we get: 
\begin{center}
    $2c_{t,s_p} \leq K\mu_p + \eta_{m(\mu_p)} - K\mu_r - \eta_{m(\mu_r)} \triangleq \Delta_{r,p}(K)$.\vspace{0.4cm}
\end{center}
Also, note that    
\begin{center}
    $2c_{t,s_p} = 2\sqrt{\frac{2\ln{t}}{s_p}} \leq \Delta_{r,p}(K), \forall r \in \mathcal{P}_{(s,d)}, r \neq p$.\vspace{0.4cm}
\end{center}
Note that we get this inequality $L - 1$ times, for all $r \neq p$. Next, we define:
\begin{equation}
\begin{array}{l}
\displaystyle
   \Delta_p^{min}(K)\triangleq \min\limits_{\substack{r \in \mathcal{P}_{(s,d)}\\r \neq p}}{\Delta_{r,p}(K)} \vspace{0.4cm}\\
   \displaystyle
   \hspace{1.5cm}=\min\limits_{\substack{r \in \mathcal{P}_{(s,d)}\\r \neq p}}{K\mu_p + \eta_{m(\mu_p)} - K\mu_r - \eta_{m(\mu_r)}}.
\end{array}
\end{equation}
Now, we choose $\tilde{s}_p\in\mathbb{R}$ such that $2c_{t,\tilde{s}_p} = |\Delta_p^{min}(K)|$ holds. Thus,\vspace{0.4cm}
\begin{center}
$\displaystyle 2\sqrt{\frac{2\ln{t}}{\tilde{s}_p}}= |\Delta_p^{min}(K)|$,
\end{center}
and we get
\begin{center}
$\displaystyle \frac{8\ln{t}}{\tilde{s}_p} = (\Delta_p^{min}(K))^2$,
\end{center}
and finally, we get
\begin{equation}
\displaystyle
\tilde{s}_p = \frac{8\ln{t}}{(\Delta_p^{min}(K))^2}.
\end{equation}
Next, recall that by definition $s_p \geq l$ and $l \in \mathbb{N}$. Then,\vspace{0.4cm}

\begin{center}
    $\left\lceil{\frac{8\ln{t}}{(\Delta_p^{min}(K))^2}}\right\rceil = l \leq s_p$.\vspace{0.4cm} 
\end{center}

Now, we can upper bound $T_p(n)$ as follows:\vspace{0.4cm}
\begin{equation*}
\begin{array}{l}
\displaystyle T_p(n) \leq \left\lceil{\frac{8\ln{t}}{(\Delta_p^{min}(K))^2}}\right\rceil \vspace{0.4cm}\\
\displaystyle
         + \sum_{t=1}^{n}\sum_{\substack{r \in \mathcal{P}_{(s,d)}\\r \neq p}}\sum_{s_g=1}^{t-1}\sum_{s_p=l}^{t-1} \bigg[Pr(K\bar{C}_r(s_g) \geq K\mu_r + c_{t,s_g})\vspace{0.4cm}\\
\displaystyle
\hspace{3.5cm}
+ Pr(K\bar{C}_{p}(s_p) \leq K\mu_p - c_{t,s_p})\bigg]         
\end{array}
\end{equation*}
\begin{equation*}
\begin{array}{l}
\displaystyle
\leq \left\lceil{\frac{8\ln{t}}{(\Delta_p^{min}(K))^2}}\right\rceil + \sum_{t=1}^{n}\sum_{\substack{r \in \mathcal{P}_{(s,d)}\\r \neq p}}\sum_{s_g=1}^{t-1}\sum_{s_p=l}^{t-1}2t^{-4}
\end{array}
\end{equation*}
\begin{equation*}
\begin{array}{l}
\displaystyle
\leq \left\lceil{\frac{8\ln{t}}{(\Delta_p^{min}(K))^2}}\right\rceil + \sum_{t=1}^{\infty}\sum_{\substack{r \in \mathcal{P}_{(s,d)}\\r \neq p}}\sum_{s_g=1}^{t}\sum_{s_p=1}^{t}2t^{-4}
\end{array}
\end{equation*}
\begin{equation*}
\begin{array}{l}
\displaystyle
= \left\lceil{\frac{8\ln{t}}{(\Delta_p^{min}(K))^2}}\right\rceil + (L - 1)\sum_{t=1}^{\infty}2t^{-2}
\end{array}
\end{equation*}
\begin{equation}
\label{eq:proof_end}
\begin{array}{l}
\displaystyle
\leq \left\lceil{\frac{8\ln{t}}{(\Delta_p^{min}(K))^2}}\right\rceil + (L - 1)(1 + \frac{\pi^2}{3}).\vspace{0.4cm}
\end{array}
\end{equation}
Finally, we substitute (\ref{eq:proof_end}) in (\ref{eq:reg_bound}), which completes the proof.\vspace{0.4cm}

\bibliographystyle{ieeetr}
\bibliography{References}

\begin{thebibliography}{10}

\bibitem{amar2021online}
O.~Amar and K.~Cohen, ``Online learning for shortest path and backpressure
  routing in wireless networks,'' in {\em IEEE International Symposium on
  Information Theory (ISIT)}, pp.~2702--2707, 2021.

\bibitem{liu2012adaptive}
K.~Liu and Q.~Zhao, ``Adaptive shortest-path routing under unknown and
  stochastically varying link states,'' in {\em 10th International Symposium on
  Modeling and Optimization in Mobile, Ad Hoc and Wireless Networks (WiOpt)},
  pp.~232--237, IEEE, 2012.

\bibitem{tehrani2013distributed}
P.~Tehrani and Q.~Zhao, ``Distributed online learning of the shortest path
  under unknown random edge weights,'' in {\em IEEE International Conference on
  Acoustics, Speech and Signal Processing (ICASSP)}, pp.~3138--3142, 2013.

\bibitem{pourpeighambar2019joint}
B.~Pourpeighambar, M.~Dehghan, and M.~Sabaei, ``Joint routing and channel
  assignment using online learning in cognitive radio networks,'' {\em Wireless
  Networks}, vol.~25, no.~5, pp.~2407--2421, 2019.

\bibitem{scarlett2019overlapping}
J.~Scarlett, I.~Bogunovic, and V.~Cevher, ``Overlapping multi-bandit best arm
  identification,'' in {\em IEEE International Symposium on Information Theory
  (ISIT)}, pp.~2544--2548, 2019.

\bibitem{zhao2020novel}
L.~Zhao, W.~Zhao, A.~Hawbani, A.~Y. Al-Dubai, G.~Min, A.~Y. Zomaya, and
  C.~Gong, ``Novel online sequential learning-based adaptive routing for edge
  software-defined vehicular networks,'' {\em IEEE Transactions on Wireless
  Communications}, vol.~20, no.~5, pp.~2991--3004, 2020.

\bibitem{salameh2020intelligent}
H.~B. Salameh, S.~Otoum, M.~Aloqaily, R.~Derbas, I.~Al~Ridhawi, and
  Y.~Jararweh, ``Intelligent jamming-aware routing in multi-hop {IoT}-based
  opportunistic cognitive radio networks,'' {\em Ad Hoc Networks}, vol.~98,
  p.~102035, 2020.

\bibitem{raj2020survey}
R.~N. Raj, A.~Nayak, and M.~S. Kumar, ``A survey and performance evaluation of
  reinforcement learning based spectrum aware routing in cognitive radio ad hoc
  networks,'' {\em International Journal of Wireless Information Networks},
  vol.~27, no.~1, pp.~144--163, 2020.

\bibitem{gafni2021federated}
T.~Gafni, N.~Shlezinger, K.~Cohen, Y.~C. Eldar, and H.~V. Poor, ``Federated
  learning: A signal processing perspective,'' {\em to appear in the IEEE
  Signal Processing Magazine, arXiv preprint arXiv:2103.17150}, 2021.

\bibitem{huang2021tsor}
Z.~Huang, Y.~Xu, and J.~Pan, ``Tsor: Thompson sampling-based opportunistic
  routing,'' {\em IEEE Transactions on Wireless Communications}, vol.~20,
  no.~11, pp.~7272--7285, 2021.

\bibitem{somekh2008cooperative}
A.~Somekh-Baruch, S.~Shamai, and S.~Verd{\'u}, ``Cooperative multiple-access
  encoding with states available at one transmitter,'' {\em IEEE Transactions
  on Information Theory}, vol.~54, no.~10, pp.~4448--4469, 2008.

\bibitem{gai2010learning}
Y.~Gai, B.~Krishnamachari, and R.~Jain, ``Learning multiuser channel
  allocations in cognitive radio networks: A combinatorial multi-armed bandit
  formulation,'' in {\em IEEE Symposium on New Frontiers in Dynamic Spectrum
  Access Networks (DySPAN)}, pp.~1--9, 2010.

\bibitem{he2013endhost}
T.~He, D.~Goeckel, R.~Raghavendra, and D.~Towsley, ``Endhost-based shortest
  path routing in dynamic networks: An online learning approach,'' in {\em
  Proceedings of the IEEE INFOCOM}, pp.~2202--2210, 2013.

\bibitem{ghosh2016secondary}
A.~Ghosh and S.~Sarkar, ``Secondary spectrum oligopoly market over large
  locations,'' in {\em IEEE Information Theory and Applications Workshop
  (ITA)}, pp.~1--10, 2016.

\bibitem{talebi2017stochastic}
M.~S. Talebi, Z.~Zou, R.~Combes, A.~Proutiere, and M.~Johansson, ``Stochastic
  online shortest path routing: The value of feedback,'' {\em IEEE Transactions
  on Automatic Control}, vol.~63, no.~4, pp.~915--930, 2017.

\bibitem{zhao2007survey}
Q.~Zhao and B.~M. Sadler, ``A survey of dynamic spectrum access,'' {\em IEEE
  signal processing magazine}, vol.~24, no.~3, pp.~79--89, 2007.

\bibitem{srikant2013communication}
R.~Srikant and L.~Ying, {\em Communication networks: an optimization, control,
  and stochastic networks perspective}.
\newblock Cambridge University Press, 2013.

\bibitem{gong2016distributed}
H.~Gong, L.~Fu, X.~Fu, L.~Zhao, K.~Wang, and X.~Wang, ``Distributed multicast
  tree construction in wireless sensor networks,'' {\em IEEE Transactions on
  Information Theory}, vol.~63, no.~1, pp.~280--296, 2016.

\bibitem{stolyar2005maximizing}
A.~L. Stolyar, ``Maximizing queueing network utility subject to stability:
  Greedy primal-dual algorithm,'' {\em Queueing Systems}, vol.~50, no.~4,
  pp.~401--457, 2005.

\bibitem{eryilmaz2006joint}
A.~Eryilmaz and R.~Srikant, ``Joint congestion control, routing, and {MAC} for
  stability and fairness in wireless networks,'' {\em IEEE Journal on Selected
  Areas in Communications}, vol.~24, no.~8, pp.~1514--1524, 2006.

\bibitem{bui2009novel}
L.~Bui, R.~Srikant, and A.~Stolyar, ``Novel architectures and algorithms for
  delay reduction in back-pressure scheduling and routing,'' in {\em IEEE
  International Conference on Computer Communications (INFOCOM)},
  pp.~2936--2940, 2009.

\bibitem{sinha2017optimal}
A.~Sinha and E.~Modiano, ``Optimal control for generalized network-flow
  problems,'' {\em IEEE/ACM Transactions on Networking}, vol.~26, no.~1,
  pp.~506--519, 2017.

\bibitem{joo2011performance}
C.~Joo, ``On the performance of back-pressure scheduling schemes with
  logarithmic weight,'' {\em IEEE transactions on wireless communications},
  vol.~10, no.~11, pp.~3632--3637, 2011.

\bibitem{ying2010combining}
L.~Ying, S.~Shakkottai, A.~Reddy, and S.~Liu, ``On combining shortest-path and
  back-pressure routing over multihop wireless networks,'' {\em IEEE/ACM
  Transactions on Networking}, vol.~19, no.~3, pp.~841--854, 2010.

\bibitem{menache2008rate}
I.~Menache and N.~Shimkin, ``Rate-based equilibria in collision channels with
  fading,'' {\em IEEE Journal on Selected Areas in Communications}, vol.~26,
  no.~7, pp.~1070--1077, 2008.

\bibitem{menache2011network}
I.~Menache and A.~Ozdaglar, ``Network games: Theory, models, and dynamics,''
  {\em Synthesis Lectures on Communication Networks}, vol.~4, no.~1,
  pp.~1--159, 2011.

\bibitem{cohen2015distributed}
K.~Cohen and A.~Leshem, ``Distributed game-theoretic optimization and
  management of multichannel {ALOHA} networks,'' {\em IEEE/ACM Transactions on
  Networking}, vol.~24, no.~3, pp.~1718--1731, 2015.

\bibitem{cohen2016distributed}
K.~Cohen, A.~Nedi{\'c}, and R.~Srikant, ``Distributed learning algorithms for
  spectrum sharing in spatial random access wireless networks,'' {\em IEEE
  Transactions on Automatic Control}, vol.~62, no.~6, pp.~2854--2869, 2017.

\bibitem{bistritz2018game}
I.~Bistritz and A.~Leshem, ``Game theoretic dynamic channel allocation for
  frequency-selective interference channels,'' {\em IEEE Transactions on
  Information Theory}, vol.~65, no.~1, pp.~330--353, 2018.

\bibitem{cao2014qos}
Y.~Cao, D.~Duan, X.~Cheng, L.~Yang, and J.~Wei, ``{QoS}-oriented wireless
  routing for smart meter data collection: Stochastic learning on graph,'' {\em
  IEEE transactions on wireless communications}, vol.~13, no.~8,
  pp.~4470--4482, 2014.

\bibitem{tekin2011online}
C.~Tekin and M.~Liu, ``Online learning in opportunistic spectrum access: A
  restless bandit approach,'' in {\em IEEE International Conference on Computer
  Communications (INFOCOM)}, pp.~2462--2470, 2011.

\bibitem{tekin2012online}
C.~Tekin and M.~Liu, ``Online learning of rested and restless bandits,'' {\em
  IEEE Transactions on Information Theory}, vol.~58, no.~8, pp.~5588--5611,
  2012.

\bibitem{liu2012learning}
H.~Liu, K.~Liu, and Q.~Zhao, ``Learning in a changing world: Restless
  multiarmed bandit with unknown dynamics,'' {\em IEEE Transactions on
  Information Theory}, vol.~59, no.~3, pp.~1902--1916, 2012.

\bibitem{cohen2014restless}
K.~Cohen, Q.~Zhao, and A.~Scaglione, ``Restless multi-armed bandits under
  time-varying activation constraints for dynamic spectrum access,'' in {\em
  48th Asilomar Conference on Signals, Systems and Computers}, pp.~1575--1578,
  IEEE, 2014.

\bibitem{gafni2018learning}
T.~Gafni and K.~Cohen, ``Learning in restless multi-armed bandits using
  adaptive arm sequencing rules,'' in {\em IEEE International Symposium on
  Information Theory (ISIT)}, pp.~1206--1210, 2018.

\bibitem{bistritz2018distributed}
I.~Bistritz and A.~Leshem, ``Distributed multi-player bandits-a game of thrones
  approach,'' in {\em Advances in Neural Information Processing Systems},
  pp.~7222--7232, 2018.

\bibitem{turgay2019exploiting}
E.~Tur{\u{g}}ay, C.~Bulucu, and C.~Tekin, ``Exploiting relevance for online
  decision-making in high-dimensions,'' {\em IEEE Transactions on Signal
  Processing}, vol.~69, pp.~1438--1451, 2020.

\bibitem{yemini2020restless}
M.~Yemini, A.~Leshem, and A.~Somekh-Baruch, ``Restless hidden markov bandit
  with linear rewards,'' in {\em IEEE Conference on Decision and Control
  (CDC)}, pp.~1183--1189, 2020.

\bibitem{gafni2020learning}
T.~Gafni and K.~Cohen, ``Learning in restless multiarmed bandits via adaptive
  arm sequencing rules,'' {\em IEEE Transactions on Automatic Control},
  vol.~66, no.~10, pp.~5029--5036, 2020.

\bibitem{gafni2021distributed}
T.~Gafni and K.~Cohen, ``Distributed learning over markovian fading channels
  for stable spectrum access,'' {\em arXiv preprint arXiv:2101.11292}, 2021.

\bibitem{gafni2021learning}
T.~Gafni, M.~Yemini, and K.~Cohen, ``Learning in restless bandits under
  exogenous global markov process,'' {\em arXiv preprint arXiv:2112.09484},
  2021.

\bibitem{agrawal1995sample}
R.~Agrawal, ``Sample mean based index policies with {$O(log n)$} regret for the
  multi-armed bandit problem,'' {\em Advances in Applied Probability},
  pp.~1054--1078, 1995.

\bibitem{auer2002finite}
P.~Auer, N.~Cesa-Bianchi, and P.~Fischer, ``Finite-time analysis of the
  multiarmed bandit problem,'' {\em Machine learning}, vol.~47, no.~2-3,
  pp.~235--256, 2002.

\bibitem{tabei2021multi}
G.~Tabei, Y.~Ito, T.~Kimura, and K.~Hirata, ``Multi-armed bandit-based routing
  method for in-network caching,'' in {\em IEEE Asia-Pacific Signal and
  Information Processing Association Conference (APSIPA)}, pp.~1899--1902,
  2021.

\bibitem{wang2018deep}
S.~Wang, H.~Liu, P.~H. Gomes, and B.~Krishnamachari, ``Deep reinforcement
  learning for dynamic multichannel access in wireless networks,'' {\em IEEE
  Transactions on Cognitive Communications and Networking}, vol.~4, no.~2,
  pp.~257--265, 2018.

\bibitem{yu2019deep}
Y.~Yu, T.~Wang, and S.~C. Liew, ``Deep-reinforcement learning multiple access
  for heterogeneous wireless networks,'' {\em IEEE Journal on Selected Areas in
  Communications}, vol.~37, no.~6, pp.~1277--1290, 2019.

\bibitem{naparstek2017deep}
O.~Naparstek and K.~Cohen, ``Deep multi-user reinforcement learning for dynamic
  spectrum access in multichannel wireless networks,'' in {\em IEEE Global
  Communications Conference (GLOBECOM)}, pp.~1--7, 2017.

\bibitem{naparstek2018deep}
O.~Naparstek and K.~Cohen, ``Deep multi-user reinforcement learning for
  distributed dynamic spectrum access,'' {\em IEEE Transactions on Wireless
  Communications}, vol.~18, no.~1, pp.~310--323, 2019.

\bibitem{lin2010autonomic}
Z.~Lin and M.~van~der Schaar, ``Autonomic and distributed joint routing and
  power control for delay-sensitive applications in multi-hop wireless
  networks,'' {\em IEEE Transactions on Wireless Communications}, vol.~10,
  no.~1, pp.~102--113, 2010.

\bibitem{tang2016joint}
F.~Tang and J.~Li, ``Joint rate adaptation, channel assignment and routing to
  maximize social welfare in multi-hop cognitive radio networks,'' {\em IEEE
  Transactions on Wireless Communications}, vol.~16, no.~4, pp.~2097--2110,
  2016.

\bibitem{rong2008enhanced}
B.~Rong, Y.~Qian, K.~Lu, and R.~Q. Hu, ``Enhanced {QoS} multicast routing in
  wireless mesh networks,'' {\em IEEE Transactions on Wireless Communications},
  vol.~7, no.~6, pp.~2119--2130, 2008.

\bibitem{awerbuch2008online}
B.~Awerbuch and R.~Kleinberg, ``Online linear optimization and adaptive
  routing,'' {\em Journal of Computer and System Sciences}, vol.~74, no.~1,
  pp.~97--114, 2008.

\bibitem{auer2002nonstochastic}
P.~Auer, N.~Cesa-Bianchi, Y.~Freund, and R.~E. Schapire, ``The nonstochastic
  multiarmed bandit problem,'' {\em SIAM journal on computing}, vol.~32, no.~1,
  pp.~48--77, 2002.

\bibitem{moeller2010routing}
S.~Moeller, A.~Sridharan, B.~Krishnamachari, and O.~Gnawali, ``Routing without
  routes: The backpressure collection protocol,'' in {\em Proceedings of the
  9th ACM/IEEE International Conference on Information Processing in Sensor
  Networks}, pp.~279--290, 2010.

\end{thebibliography}

\end{document}